\documentclass[aps,twocolumn,floatfix]{revtex4-1} 

\usepackage{amsmath,amsfonts,amssymb,graphicx,color,times,bbm}

\newcommand{\zz}{{\mathbbm{Z}}}

\newcommand{\id}{{\mathbbm{1}}}
\renewcommand{\vec}[1]{\boldsymbol{#1}}
\newcommand{\me}{\mathrm{e}}
\newcommand{\mi}{\mathrm{i}}
\newcommand{\md}{\mathrm{d}}

\definecolor{green}{rgb}{0.2,.7,0.4}

\begin{document}

\title{Probing local relaxation of cold 
atoms in optical superlattices}

\author{A.\ Flesch$^1$, M.\ Cramer$^{2}$, I.\ P.\ McCulloch$^3$,  
U.\ Schollw{\"o}ck$^1$, and J.\ Eisert$^{2,4}$} 

\affiliation{
1 Institut f{\"u}r Theoretische Physik C, 
RWTH Aachen University, 52056 Aachen, Germany\\
2 Institute for Mathematical Sciences, Imperial College London,
SW7 2PE London, UK\\
3 School of Physical Sciences, The University of 
Queensland, Brisbane, QLD 4072, Australia\\
4 Institute for Physics and Astronomy, University of Potsdam,
14476 Potsdam, Germany}

\begin{abstract}
In the study of relaxation processes in coherent non-equilibrium dynamics of quenched quantum systems, 
ultracold atoms in optical superlattices with periodicity $2$ provide a very fruitful test ground. In this work, we consider the dynamics of a particular, experimentally accessible initial state prepared in a 
superlattice structure evolving under a Bose-Hubbard Hamiltonian in the entire range of interaction strengths, further investigating the issues raised in 
Ref.\ [Phys.\ Rev.\ Lett.\ {\bf 101}, 063001 (2008)]. 
We investigate the relaxation dynamics analytically in the non
interacting and hard core bosonic limits, deriving explicit expressions for the dynamics of certain correlation functions, and numerically for finite interaction strengths using the time-dependent density-matrix renormalization (t-DMRG) approach. We can identify signatures of local relaxation that can be accessed experimentally with present technology.
While the global system preserves the information
about the initial condition, locally the system relaxes to the 
state having maximum entropy respecting the constraints of the initial 
condition. For finite interaction strengths and finite times, the 
relaxation dynamics contains signatures of the relaxation 
dynamics of both the non-interacting and hard core bosonic limits. 
\end{abstract}

\maketitle

\date{\today}

\section{Introduction}

Both in classical and quantum physics, a complete framework for the description of equilibrium properties of arbitrary physical many-body systems exists, although the explicit calculation of many-body equilibrium properties is an often unsolved challenge. The situation is much less satisfactory when it comes to the study of the 
non-equilibrium properties of many-body systems, where such a general framework is missing and may not even exist.  
Research in this field has therefore focussed on relatively specific issues and types of non-equilibrium. One of the issues taking center stage is whether quantum many-body systems in non-equilibrium evolving coherently under a local Hamiltonian equilibrate or not. If so, one may ask whether the equilibrium states can in some sense be described by a thermal state. This old and fundamental question of the equilibration of quantum many-body systems has 
enjoyed quite a renaissance recently [1--36].

A specific setting of coherent non-equilibrium quantum dynamics is provided by quantum {\em quenches}, where one starts from an eigenstate of some initial Hamiltonian and pushes the system out of equilibrium 
by a sudden change (or ``quench") 
of system parameters, leading to a new Hamiltonian. One then considers the evolution of the system under the new Hamiltonian. A further restriction is provided by the assumption that the quantum system under consideration is {\em closed}, 
i.e.,\ has no coupling to a bath of degrees of freedom that might assist the relaxation process.
Time evolution (and hence the potential relaxation to some equilibrium) will obviously be constrained by the constants of motion, i.e.,\ Hermitian operators commuting with the new Hamiltonian whose expectation values are fixed by the initial state; in that sense, any relaxed state will to a certain degree show some memory of the initial state. 

It has been conjectured that---in some sense---
the quantum system should relax to the maximum entropy 
state consistent with the expectation values of the 
constants of motion fixed 
by the initial state (see, e.g., Refs.\ \cite{Rigol,Olshanii,Cazalilla}), 
also referred to as a  
generalized Gibbs ensemble \cite{Jaynes}.
This may be attractive because it reminds of Jaynes' derivation of 
equilibrium statistical mechanics.

This observation is in conflict with the fact that obviously, if the system 
can be meaningfully treated as a {\em closed}
quantum system, one cannot expect the whole
system to relax: Initially pure states will remain so in time 
under a unitary time evolution [2--6]. 
After all, the entire information about the initial condition 
is still stored in the system, albeit in a dilute fashion. 
Yet, this observation is by no means in contradiction with the 
possibility that in any local observation, the system may appear
perfectly relaxed, even without invoking a time 
average \cite{Relax,Tegmark,Barthel,Kollath}.
The key point is that {\it locally}, 
one may well expect the relaxation to be true \cite{Momentum}:
For  any subset of sites in a sufficiently large lattice system,
the reduced state may well converge to the reduced state of
the maximum entropy state, given the conserved
quantities of motion, and stay relaxed for an arbitrary
long time. Indeed, for such a subset it is, under suitable assumptions 
about its interactions with the rest of the world, easy to make contact to 
Jaynes' formulation of statistical mechanics.

Ref.\ \cite{Relax} considers a variant of the question in which
this local relaxation of subsystems, referred to as
{\it local relaxation conjecture}, can in fact be rigorously 
proven to hold exactly: This is the one where one evolves 
a state deep in a Mott phase according to a Bose-Hubbard
Hamiltonian corresponding to the deep superfluid phase---treated 
as a non-interacting system. In this setting, 
the local relaxation is in fact true: The reduced state
of a block $I$ of consecutive sites indeed converges 
to a maximum entropy state
\begin{equation}
	\hat \varrho_I(t)\rightarrow \hat\varrho_{\text{max}}
\end{equation}
(in trace norm) for large systems and large times, 
having maximum entropy consistent with 
the constants of motion \cite{Relax}. 
Note that there is no time 
average and the initial state is not a 
Gaussian state. 

Also, for free bosonic and fermionic systems, and for Gaussian
initial states, it has been shown rigorously in Refs.\
\cite{Tegmark,Barthel} under which conditions the local relaxation 
conjecture is true. It turns out that in these cases, the presence and speed of 
local relaxation depends crucially on the single-particle spectrum and 
the dimensionality of the systems.

The in some sense inverse case of a quench from the superfluid 
phase to the Mott insulating phase, but generally at finite interaction 
strengths, has been studied in Ref.\
\cite{Kollath}. In this non-integrable system, numerically two 
distinct non-equilibrium regimes have been 
found where equilibrated states resembled thermal 
states or states with memory.

The physical intuition why local relaxation happens is the
following: If one switches to a new Hamiltonian, the system
is no longer in equilibrium. Hence, one has local excitations
at each point [2, 3, 12--16].
They cannot travel arbitrarily fast, however,
as there is a finite speed of information transfer
in lattice systems \cite{LR,LRC}. At each site, in time more and
more ``waves'' of farther and farther away sites can possibly
have a significant influence on this site. This is related
to a finite speed of sound or of 
information transfer in a lattice system \cite{LRC}. 
Then, local relaxation may be a consequence of the  
incommensurate influence
of these excitations generating mixing and thermalization.
The ``thermalization" time scale is hence governed by the speed of information 
transfer \cite{Note}.

This also links to kinematical approaches to the problem 
[37, 47--51],
arguing that most states anyway look locally very much 
relaxed in that they have large entropy. A random pure state
(as taken from the Haar measure) will locally have a large entropy. 
Specifically, in interacting systems one should expect such a local 
relaxation to be true, too, an aspect that will be studied in this work.

So far the discussion has not taken into 
account the recurrence happening in any closed quantum 
or classical system \cite{Relax}. 
For finite, but large system sizes, recurrence
times will however become so long that recurrence 
effects become negligible, 
in that the quantum many-body system can be effectively locally equilibrated on a much 
shorter time scale than the recurrence time. 

We will see that while settings exhibiting 
local relaxation may be generated in various fashions, it is 
a greater challenge to actually probe signatures of local
relaxation. This apparent dilemma---that to demonstrate
local relaxation appears to necessitate local addressing---will 
be resolved in this work, 
by making use of a period-$2$ setting, further developing
the idea of Ref.\ \cite{Letter}.
The taken path opens up a way to experimentally explore local
relaxation effects using atoms in optical superlattices.
We systematically investigate various 
aspects of local relaxation in such a setting, both
using analytical as well as numerical methods, based on
a time-dependent density-matrix 
renormalization-group (t-DMRG) approach.

\begin{figure}
\includegraphics[width=7cm]{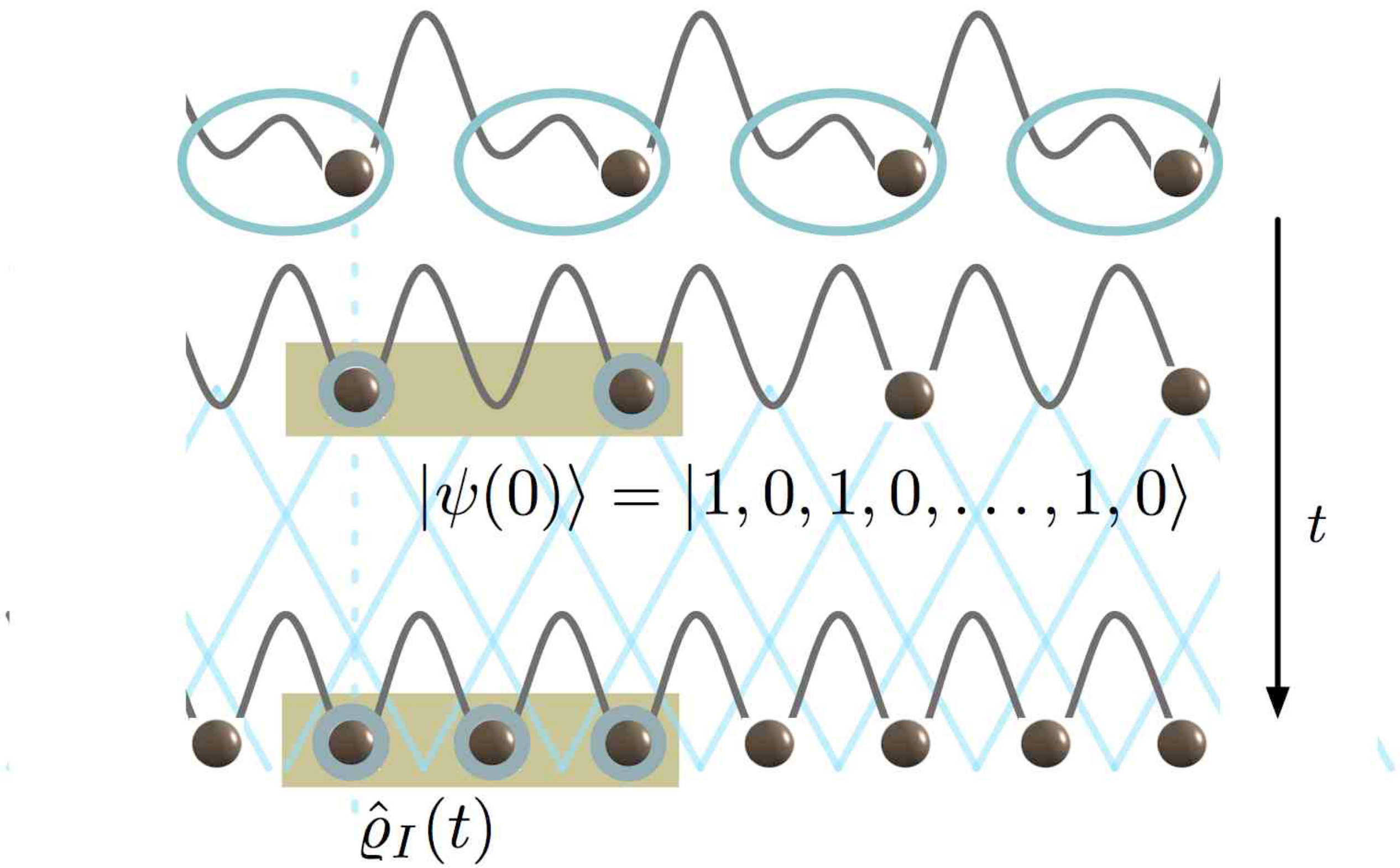}
\caption{\label{sketch}
Idealized sketch of the effect of local relaxation
in a setting having periodicity $2$. Starting from
an idealized initial condition of bosonic atoms being
present or absent, $|\psi(0)\rangle= |1,0,\dots, 1,0\rangle$, 
in even and odd sites of a one-dimensional Bose-Hubbard 
system--achieved by imposing a superlattice to the one-dimensional 
optical lattice--the system locally relaxes to an apparent
maximum entropy state.}
\end{figure}

\section{Experimental setup: Ultracold atoms in optical superlattices}

The current surge of interest in the relaxation of quantum systems after a 
quench is mainly motivated by the advent of ultracold atoms in 
optical lattices [52--56].
These systems are highly attractive as they allow for sudden controlled 
manipulations of system parameters, hence quenches, 
are strongly interacting, hence non-trivial, are on experimental time scales 
essentially closed quantum systems,
and therefore show coherent quantum dynamics. Systems are 
also sufficiently large to show non-trivial many-body behavior.

In the present context, however, the major drawback of ultracold atoms is 
that despite the unprecedented possibilities of manipulation the study of 
{\em local relaxation} provides an experimental challenge. This is due to the 
fact that local (i.e.,\ site-resolved) measurements on ultracold atoms in optical lattices are still not 
satisfactory, albeit rapid progress is being made.
In this work, we propose to study instances of local 
relaxation even in a setting where strictly local quantities can not easily
be studied by using {\em optical superlattices} 
[57--60].

Following the setup very recently  realized by Bloch and coworkers 
\cite{Wells,TimeResolved}, 
we consider bosonic Rubidium-87 atoms in a period-$2$ optical superlattice geometry: 
Two standing-wave laser fields at wavelengths $765$nm and $1530$nm are 
superimposed with fixed relative location to provide a superlattice geometry of 
lattice constants $a=382.5$nm and $2a=765$nm. It is experimentally possible, 
among other things, to (i) change the relative strength of the two optical lattices 
and (ii) shift their relative position, by altering phase and detuning of 
the optical superlattice. 
Assuming the strength of the lattice with lattice constant $a$ to be fixed, 
(i) allows to couple and decouple this lattice into an array of double well 
potentials, combining one {\em odd} (o) and one {\em even} (e) site of 
the original lattice. In such double well potentials, (ii) allows to introduce a bias 
between the chemical potentials of the odd and even sites, $\Delta = \mu_o - \mu_e$.  

Isolating double wells and biasing odd vs.\ even sites, in turn, allows for the 
preparation of {\em patterns} of atoms and for extracting {\em local} quantities 
to the degree that odd and even sites can be distinguished from each other:
\begin{itemize}
\item {\em Preparation of periodic patterns} is achieved by loading 
the superlattice while introducing a bias between odd and even 
sites such that due to a shifting of particles all particles are on either 
odd or even sites after loading. Using further experimental techniques 
\cite{Wells}, multiple occupancies on the occupied sites can be 
eliminated, leaving a sequence of empty and single-occupied sites.

\item {\em Period-$2$ local} measurements can be obtained by mapping 
odd and even sites to different Brillouin zones: Assuming completely 
decoupled double wells, each part of the double well has multiple bands 
separated by well-defined energies. Biasing, say, the odd sites relative to 
the even sites by an energy in excess of the separation energy of the 
band-separation energy, odd-site particles are reloaded into the higher 
band of the even sites, whereas the even-site particles stay in the lower band. 
A standard time-of-flight mapping then shows the even-site particles in the 
first Brillouin zone, the odd-site particles in the higher  Brillouin zones.

\item {\em More sophisticated measurements with period-$2$} can also 
be performed in principle.
When letting the system evolve for some defined hold time, and then freezing
the time evolution by ramping up the barrier, one can also measure 
nearest-neighbor correlations in the lattice
\cite{Wells,TimeResolved,Superlattice,Superlattice2}. 
There are several ways to proceed here: On the one hand, one can isolate 
double wells, let them dephase. Then upon free expansion,
the average correlations can be measured as in a double
slit experiment. One the other hand, correlations can be
mapped to densities.  
To the extent that the barrier between the wells is
sufficiently high such that the time evolution can be described
by a collection of double well systems (and higher Wannier bands do
not have to be taken into account), we can investigate the time
evolution in each individual double well. Up to the presence of a
confining potential, the total time evolution then reflects the time 
evolution in each of the
double wells. Let us label the two modes in any of the wells by
$1$ and $2$, one may apply an appropriate
free Hamiltonian to map correlations onto on-site densities.
Specifically, for $U=0$ we find that
\begin{equation}\label{map}
2 \, {\text{im}} \langle \hat{a}_1^\dagger \hat{a}_2\rangle
	 =\langle \hat{a}_2^\dagger (t)\hat{a}_2 (t)\rangle-
	\langle \hat{a}_1^\dagger (t)\hat{a}_1 (t)\rangle),
\end{equation}
when letting $\hat{a}_1,\hat{a}_2$ evolve under the free Hamiltonian
\begin{equation}	
	\hat h=-\left(\hat{a}_1^\dagger \hat{a}_2  + \hat{a}_2^\dagger \hat{a}_1\right),
\end{equation}
until $t=\pi/4$. One can hence measure
the imaginary part of the appropriate correlators, and hence
the period-$2$ correlators in the full lattice. In settings where in
the Heisenberg picture the phase map $\hat a_2\mapsto i \hat a_2$ for each
right well is feasible, one can also measure the real part. If one has
small interactions (because of the Wannier band problem---a description
in terms of Wannier functions still has to be valid, accompanied
by a non-vanishing $U$), then this 
leads to a small dephasing in
this mapping. Direct numerical simulation can take this properly into 
account, allowing for a correct interpretation of the 
experimental observation. Such a technique allows to measure 
correlators from a time-resolved
observation of on-site densities. Similar
time-resolved measurements have 
recently been performed in 
optical superlattices \cite{TimeResolved}. 

\end{itemize}

As shown in Fig.\ \ref{sketch}, we propose to start from a two-periodic initial state 
prepared by the superlattice setup as described above, where all odd sites are occupied by exactly 1 boson and all even sites are empty. The initial
state vector of the entire lattice is hence 
\begin{equation}
	|\psi(0)\rangle=|1,0,1,0,\dots1,0\rangle.
\end{equation}
Tools how to experimentally achieve that sites
are to a good approximation occupied by a single
atom, and not two or more atoms, are described in
Refs.\ [57--59].
If the $2a$-lattice is suddenly
switched off, i.e.,\ the system quenched, the 
state vector  will evolve in time $|\psi(t)\rangle=\me^{-\mi t\hat H/\hbar} 
|\psi(0)\rangle$
according to the 
conventional Bose-Hubbard Hamiltonian
\begin{equation}
	\hat{H}=-J\sum_{i=1}^L
	(\hat{b}^\dagger_{i+1}\hat{b}_i+ 
	\hat{b}^\dagger_{i}\hat{b}_{i+1}) +
	\frac{U}{2}\sum_{i=1}^L\hat{n}_i\left(\hat{n}_i-1\right),
\end{equation} 
where $U$ and $J$ are the standard interaction 
and hopping parameters of the Bose-Hubbard model 
that can be calculated microscopically from the lattice 
parameters. The system size is given by $L$, whereas
the boson number is $N=L/2$ from this setup. $L$ will always be even in the following, in line with the proposed setup; various boundary conditions will be imposed. We work in units where $J=1$ and $\hbar=1$.

We will not consider the effect of the occupation of higher order bands,
but will stay within the limit of applicability of the Bose-Hubbard model. 
We will for the purposes of the present paper also neither consider an
additional harmonic confining potential---except when explicitly
stated otherwise---nor additional dephasing effects due to 
statistical fluctuations of local fields. Instead, we will systematically
flesh out what behavior is expected in this slightly idealized setting,
to see how local relaxation manifests itself here, 
to comment on the impact of imperfections later.

As $\hat\varrho(0)= 
|\psi(0)\rangle\langle \psi(0)|$ is not an eigenstate 
of $\hat{H}$, we are 
in a non-equilibrium situation.
The initial state is uncorrelated and shows inhomogeneous density of 
periodicity $2$. Coherent quantum dynamics is now expected to 
homogenize densities locally 
and to build up non-local correlations.

Local relaxation is now monitored by the global measurement  of the total occupation of even, $\langle\hat{N}_e(t)\rangle$, and odd, $\langle\hat{N}_o(t)\rangle$, sites. In a translationally invariant setting, this gives access to local observables as
\begin{equation}
	\langle\hat{N}_e(t) \rangle=
	\sum_{i=1}^{L/2}\bigl\langle\hat{n}_{2i}(t)\bigr\rangle
	=\frac{L}{2}\langle\hat{n}_{2i}(t)\rangle ,
\end{equation}
and similarly for odd sites.

If one has experimental access to the variances
$\sigma^2_{tot,e,o}=\langle\hat{N}_{tot,e,o}^2\rangle-\langle\hat{N}_{tot,e,o}\rangle^2$,
density-density correlations (see also Ref.\ \cite{NoiseNoise}) 
between all even and all odd sites 
may be obtained through 
\begin{eqnarray}\nonumber
&&\frac{\sigma^2_{tot}-\sigma^2_{e}(t)-\sigma^2_{o}(t)}{2}
=\langle\hat{N}_e(t)\hat{N}_o(t)\rangle-\langle\hat{N}_e(t)\rangle\langle\hat{N}_o(t)\rangle\\
&=&\sum_{i,j=1}^{L/2}\left(\langle \hat{n}_{2i}(t)\hat{n}_{2j-1}(t)\rangle
-\langle \hat{n}_{2i}(t)\rangle\langle\hat{n}_{2j-1}(t)\rangle\right).
\label{varianceeq}
\end{eqnarray}
In fact, the value of $\hat{N}_{tot}$ is fixed to $L/2$ for the proposed pattern, 
so in repeated experiments of same length, this quantity will have 
variance zero, $\sigma_{tot}^2=0$.

Moreover, the quasi-momentum distribution
defined as
\begin{equation}
	S(q,t) = \frac{1}{L} \sum_{i,j=1}^L 
	\me^{\mi q(i-j)} \left\langle \hat{b}^\dagger_i(t) \hat
	b_j (t) \right\rangle,\;\;\;
	q\in [0,2\pi ],
\end{equation}
is also accessible as a global quantity, from 
time-of-flight measurements.

We now proceed as follows. In order to obtain analytical results 
we set $U=0$ and $U=\infty$ respectively, which are both essentially 
non-interacting limiting cases, can be solved exactly and show very 
similar, but not generally identical results (compare also
Ref.\ \cite{LightCone}). These results will 
be extended to the interacting finite-$U$ case which will be 
studied using t-DMRG method.  
For the time-scales considered it is an effectively 
quasi-exact method.

\section{Exact solutions for limiting cases}

Both limiting cases $U=0$ and $U=\infty$ are or can be mapped to free models. 
This case has been considered for Gaussian 
initial conditions in Refs.\ \cite{Tegmark,Barthel}, 
as well as for non-Gaussian product initial conditions in
Ref.\ \cite{Relax}. In these cases, the  reduced state of 
a subsystem $I$ 
\begin{equation}
	\hat{\varrho}_I (t)= {\text{tr}}_{L\backslash I}  [\hat{\varrho} (t)]
	\rightarrow \hat{\varrho}_{\text{max}}.
\end{equation}
will converge for large systems and long times to $\hat{\varrho}_{\text{max}}$, being the 
maximum entropy density operator as constrained by the initial conditions
(or, more precisely, if recurrences are considered, this will be 
true in 1-norm to an arbitrarily small approximation error 
for an arbitrarily long time). Also, fermionic Gaussian initial
states have been considered in Refs.\ \cite{Relax,Barthel}.
 Local relaxation will therefore happen in these cases. 

The physical mechanisms behind the relaxation processes in 
the cases of $U=0$ and $U=\infty$
is sightly different, however: 
In the case $U=0$, it is due to dephasing in the sense that freely 
propagating bosons lead to reduced density operator contributions of 
quickly oscillating phases that average out. In Ref.\ \cite{Relax}, 
this intuition leads even for non-Gaussian
initial states to maximum entropy states, by invoking a quantum version
of a central limit theorem, and exploiting the finite speed of information
transfer \cite{LRC}.

In the case $U=\infty$, real scattering processes happen, 
albeit of a very specific form that allows for a formal mapping to a 
non-interacting fermionic problem (and to the XX model, compare
also Ref.\ \cite{LightCone}):
The scattering is simply relegated to the internal sign structure 
of the wave function. We will also see that the two relaxation 
processes lead to different time-evolutions of 
most physical observables. 

For all cases of nonzero finite 
$U$, the situation should be somehow intermediate. 
For the specific setup chosen here, interacting particles will learn of each 
others existence only after some initial time they need to come into contact. 
We would therefore expect that for very short time scales, observables should 
evolve as in the $U=0$ limit, with a crossover in behavior for longer time 
scales when they start interacting. The question is whether for 
which interaction strengths the limiting pictures remain essentially 
valid and whether there is a genuinely different intermediate 
interaction regime with different relaxation behavior. Compared to 
Refs.\ \cite{Relax,Tegmark,Barthel}, in this work we 
focus to a lesser extent 
on rigorous mathematical convergence statements---like
invoking quantum versions of central limit theorems---but instead
put more emphasis on the phenomenology of the 
physical relaxation process as such in the $2$-periodic
setting. 

\subsection{Non-interacting bosons: $\mathbf{U=0}$}

In the translationally invariant case of non-interacting bosons, the 
Hamiltonian can be diagonalized by Fourier transforming to new 
bosonic operators $\hat{a}_1,\dots, \hat{a}_L$ to yield 
\begin{equation}
	\hat{H}_{U=0} = 
	\sum_{k=1}^L \lambda_k \hat{a}^\dagger_k \hat{a}_k,
\end{equation}
where
\begin{equation}\label{eq:lambda}
	\lambda_k=-2\cos(2\pi k/L)
\end{equation}
are the eigenvalues of the circulant 
Hamilton matrix $H$ with entries 
$H_{i,j}=-\delta_{\text{dist}(i,j),1}$.
While the real experiment will not have periodic boundary conditions, our 
calculations show that for realistic system sizes and time scales the difference 
between open and periodic boundary conditions is negligible.

In the Heisenberg picture, the initial state vector $|\psi(0)\rangle=|1,0,\dots,1,0\rangle$
remains time-independent, whereas the operators evolve in time as
\begin{equation}
\label{eq:evolution_boson}
\begin{split}
	\hat{b}_i (t) &= \sum_{j=1}^L V_{i,j}(t)\hat{b}_j (0),\;\;\; V(t)=\me^{-\mi tH},\\
	 V_{i,j}(t) &=\frac{1}{L}\sum_{k=1}^L\me^{-\mi t\lambda_k} 
	\me^{
	 2\pi\mi k (i-j)/L } .
	 \end{split}
\end{equation}
Straightforward algebra then yields the exact time-evolution of
two-point correlations, see Appendix,
\begin{equation}
\label{eq:fij}
\begin{split}
f_{i,j}&=
\left\langle \hat{b}_i^\dagger(t) \hat{b}_j(t) \right\rangle\\
&=\frac{1}{2}\delta_{i,j}-\frac{(-1)^i}{2}V_{j,i}(2t)
.
\end{split}
\end{equation}
In the thermodynamic limit $L\rightarrow \infty$ this turns into
\begin{equation}
f_{i,j}
=\frac{1}{2}\delta_{i,j}-\frac{(-1)^i\mi^{j-i}}{2}J_{j-i} (4Jt),
\end{equation}
where $J_n$ is a Bessel function of the first kind. 

From the above expression, one can derive the quasi-momentum distribution 
for finite $L$, see Appendix. For 
$Lq/(2\pi)\in\{1,\dots,L\}$ it is constant, $S(q,t)=1/2$, while one finds
\begin{equation}
S(q,t)=\frac{1}{2}+\frac{\mi}{L^2}\sum_{p=1}^L\me^{4J\mi t\cos(2\pi p/L)}\frac{\sin^2(Lq/2)}{\sin(2\pi p/L-q)}.
\end{equation}
for all other $q\in[0,2\pi]$.

With slightly more effort, density-density correlations as
in Eq.\ (\ref{varianceeq}) 
emerge as (see Appendix)
\begin{eqnarray}
\label{dd}
\langle \hat{n}_{2i}(t)\hat{n}_{2j-1}(t)\rangle
	-\langle \hat{n}_{2i}(t)\rangle\langle\hat{n}_{2j-1}(t)\rangle\hspace{2cm}\\
	=
	f_{2i,2j-1}f_{2j-1,2i}
-2\sum_{k=1}^{L/2}|V_{2i,2k-1}(t)V_{2j-1,2k-1}(t)|^2,\nonumber
\end{eqnarray}
which, as they are local quantities, relax for large systems and long times. For large $L$ one finds for the
global density-density correlator (see Appendix)
\begin{equation}
\begin{split}
&\langle\hat{N}_e(t)\hat{N}_o(t)\rangle-\langle\hat{N}_e(t)\rangle\langle\hat{N}_o(t)\rangle\\
&\hspace{0.4cm}\rightarrow
-\frac{L}{16}\left(
3+
J_0(8Jt)
-4[J_0(4Jt)]^2\right).
\end{split}
\end{equation}

We also show in detail
the exact local relaxation to a maximum entropy state for the case of the
initial condition $|\psi(0)\rangle = |1,0,\dots, 1,0\rangle$, largely following
Ref.\ \cite{Relax}: For every block of consecutive sites $I=\{1,\dots, s\}$,
every small
error $\varepsilon>0$ and any desired, arbitrarily long
recurrence time $t_{\text{rec}}>0$ there exists
a system size $L$ and a local relaxation time $t_{\text{rel}}>0$
such that
\begin{equation}
	\| \hat \varrho_I(t) - \hat \varrho_{\text{max}}\|_1<\varepsilon
\end{equation}
for all times $t\in [t_{\text{rel}}, t_{\text{rec}}]$, see Appendix. Here, $\|\cdot \|_1$
denotes the trace norm. Hence, after the quench for $U=0$, for $2$-periodic
initial conditions, the local state becomes in time a maximum
entropy state, to an arbitrarily good approximation.

\subsection{Hardcore bosons: $\mathbf{U\rightarrow\infty}$}

In the limit $U\rightarrow\infty$, the interaction manifests 
itself exclusively in that bosonic occupation numbers are upper bounded by $1$. 
This leads to a well-known mapping in case of quantum
ground states: The hard core limit is equivalent to the XX spin model and a model of
spinless free
fermions. But even in time evolution, in the limit of large $U$, the 
population of sites with particle number larger than unity is dynamically
suppressed to an arbitrary extent: The expectation value of the new
Hamiltonian is obviously preserved under time evolution, 
$\text{tr}[\hat H \hat\varrho(0)]=\text{tr}[\hat H \hat\varrho(t)]$,
$H=- \hat H_0+U \hat H_1$, where 
\begin{equation}
	\hat{H}_0=\sum_{i=1}^L
	(\hat{b}^\dagger_{i+1}\hat{b}_i+ 
	\hat{b}^\dagger_{i}\hat{b}_{i+1}),\,
	\hat H_1=
	\frac{1}{2}\sum_{i=1}^L\hat{n}_i\left(\hat{n}_i-1\right).
\end{equation} 
For initial states that are supported on span$(|0\rangle, |1\rangle)$ one has  $\text{tr}[\hat H \hat\varrho(0)]=-\text{tr}[\hat H_0 \hat\varrho(0)]$. Furthermore $\text{tr}[\hat H \hat\varrho(t)]\geq 
\text{tr}[(-\hat H_0+U \hat H_2)\hat\varrho(t)]$,
where 
\begin{equation}
	\hat H_2= \sum_{i=1}^L \sum_{k=2}^\infty |k\rangle\langle k|,
\end{equation}
where the latter projector defined on each site $i$. This, in turn, means that
\begin{equation}
\frac{\text{tr}[\hat H_0(\hat\varrho(t)-\hat\varrho(0))]}{U}\ge
\text{tr}[ \hat H_2\hat\varrho(t)]\ge 0.
\end{equation}
Hence, one can ensure to arbitrary accuracy for large $U$ that $\hat\varrho(t)$ is 
only supported on $\text{span}(|0\rangle,|1\rangle)$.
One therefore 
arrives in a perfectly meaningful way at the familiar hard core limit of the
Bose-Hubbard model, which is equivalent to a free fermionic spinless
system.

This mapping to non-interacting spinless 
fermions is done through the familiar Jordan-Wigner transformation
\begin{equation}
\label{eq:jordan}
\hat{b}_n = \me^{-\mi \pi \sum_{m<n} \hat{c}^\dagger_m \hat{c}_m} \hat{c}_n .
\end{equation}
Under this transformation the initial state vectors 
turn into
\begin{equation}
\label{eq:state_trans}
	| \psi \rangle =   \hat{c}_1^\dagger(0)  
	\hat{c}_3^\dagger(0)  \hat{c}_5^\dagger(0) 	
	\ldots | 0 \rangle 
\end{equation}
and the Hamiltonian reads upon a Fourier transformation to new 
fermionic operators $\hat{d}_1,\dots,\hat{d_L}$ 
\begin{equation}
	\hat{H}_{U=\infty} = 
	\sum_{k=1}^L \lambda_k \hat{d}^\dagger_k \hat{d}_k 
\end{equation}
with $\lambda_k$ as in Eq.\ (\ref{eq:lambda}) and the time-evolution of the operators is given by
\begin{equation}
\label{eq:evolution_fermion}
\hat{c}_i (t) = \sum_{j=1}^L V_{i,j}(t) \hat{c}_j(0)
\end{equation}
with $V_{i,j}(t)$ as in Eq.\ (\ref{eq:evolution_boson}).
This is formally identical to the $U=0$ case. However, there are two differences: 
Periodic boundary conditions in the original bosonic model map to (anti-)periodic boundary conditions
$\hat{c}^\dagger_{L+1} = (-1)^{N+1} \hat{c}^\dagger_{1} $.
Hence, boundary conditions stay periodic if the particle number $N$ is odd, to which we 
restrict ourselves in the following. 

The second, more important difference is that differences in results will show up due to the Jordan-Wigner transformation of operators and the difference between bosonic and fermionic (anti)commutators. It is easily shown that local densities 
and correlations between nearest neighbors translate directly (see \ref{appendix:analytics:infinity}), 
such that the respective results for $U=\infty$ are identical to those for $U=0$. 
For longer-ranged two-point correlations, including structure functions, as well as 
density-density correlations results differ. Density-density correlations are given by 
(see \ref{appendix:analytics:infinity})
\begin{equation}
\label{eq:dd_infty}
\begin{split}
\langle \hat{n}_{2i}(t)\hat{n}_{2j-1}(t)\rangle
	-\langle \hat{n}_{2i}(t)\rangle\langle\hat{n}_{2j-1}(t)\rangle\hspace{2cm}\\
	=
	-f_{2i,2j-1}f_{2j-1,2i},
\end{split}
\end{equation}
where $f_{i,j}$ is as in Eq.\ (\ref{eq:fij}). 
For large $L$ the global density-density correlator is then given by (see Appendix)
\begin{equation}
\begin{split}
&\langle\hat{N}_e(t)\hat{N}_o(t)\rangle-\langle\hat{N}_e(t)\rangle\langle\hat{N}_o(t)\rangle\\
&\rightarrow - \frac{L}{16} \left(1-J_0(8Jt)\right).
\end{split}
\end{equation}

\subsection{Interacting softcore bosons: Finite $\mathbf{U}$}

In this case, we are no longer facing an integrable model. In order
to study the relaxation dynamics, we will therefore 
turn to the time-dependent variant \cite{VidalTEBD,Daley,WhiteFeiguin} 
of the DMRG (density-matrix renormalization group) method \cite{White,RMP}. 
This method allows to follow the coherent time-evolution of strongly interacting 
quantum systems very precisely. Its reach in time is, however, 
limited by the growth of entanglement in quantum systems: 
Linear entanglement growth, for example, leads to an exponential 
growth in numerical resources needed. As we will see, interestingly,
the system under study is characterized by very strong linear entanglement 
growth. In the free instances this linear increase
is indeed provably correct \cite{Relax,Schuch}, see Sec.\ \ref{ED}.
Hence, the reachable times are quite short 
($J t \approx  6$) with up to $5000$ states kept in the simulations. 
For most quantities, this turns out to be sufficiently long to make 
contact to the non-interacting results and to read off long-time behavior. 
In particular, the results allow for a quantitative comparison to 
experiments. 

A subtlety arises from the fact that DMRG prefers open boundary conditions, 
and is hence closer to experiment. However, we would like also to 
compare to analytical results, where periodic boundary conditions 
are preferable. As it will turn out, for system sizes and times considered, 
the difference is negligible.

\section{Time evolution of densities}

For the time evolution of local densities both exactly solvable cases $U=0$ and $U=\infty$ give the same result
\begin{equation}
\begin{split}
\left\langle \hat{n}_i (t)  \right \rangle &=\frac{1}{2}-\frac{(-1)^i}{2L}\sum_{k=1}^L\me^{4\mi tJ\cos(2\pi k/L)} \\ 
&\rightarrow \frac{1}{2}-\frac{(-1)^i}{2}J_{0} (4Jt) \quad (L\rightarrow\infty) .
\end{split}
\end{equation}
Odd- and even-site densities relax symmetrically about the $n=1/2$ axis to $n=1/2$, with an asymptotic decay as $ t^{-1/2}+ o( t^{-1/2})$.

\begin{figure}
\includegraphics[width=8cm]{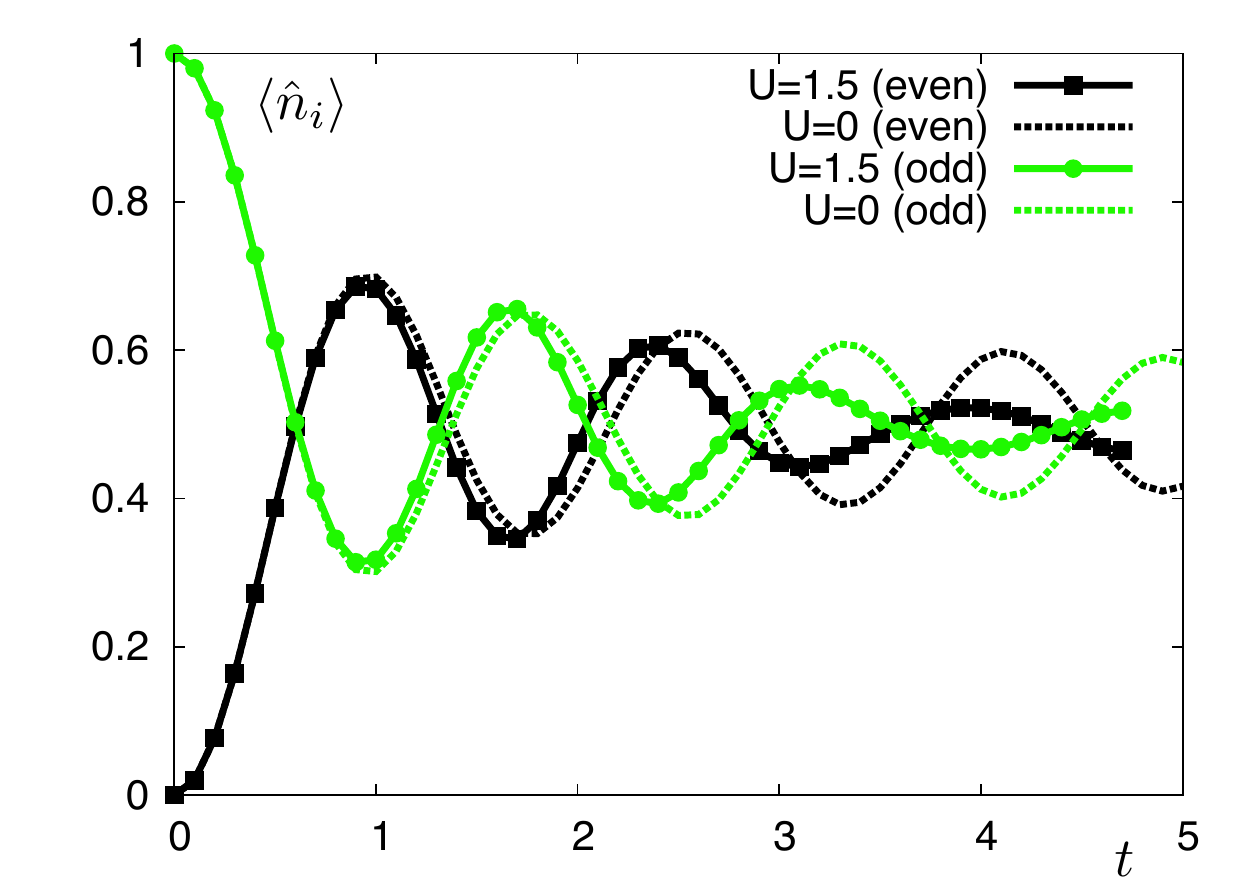}
\caption{\label{F1}Local density $\langle \hat{n}_i (t) \rangle$ vs.\ time, showing local relaxation. Shown is the time evolution of an even and an odd site for $U=0$ and $U=1.5$.}
\end{figure}

\begin{figure}
\includegraphics[width=8cm]{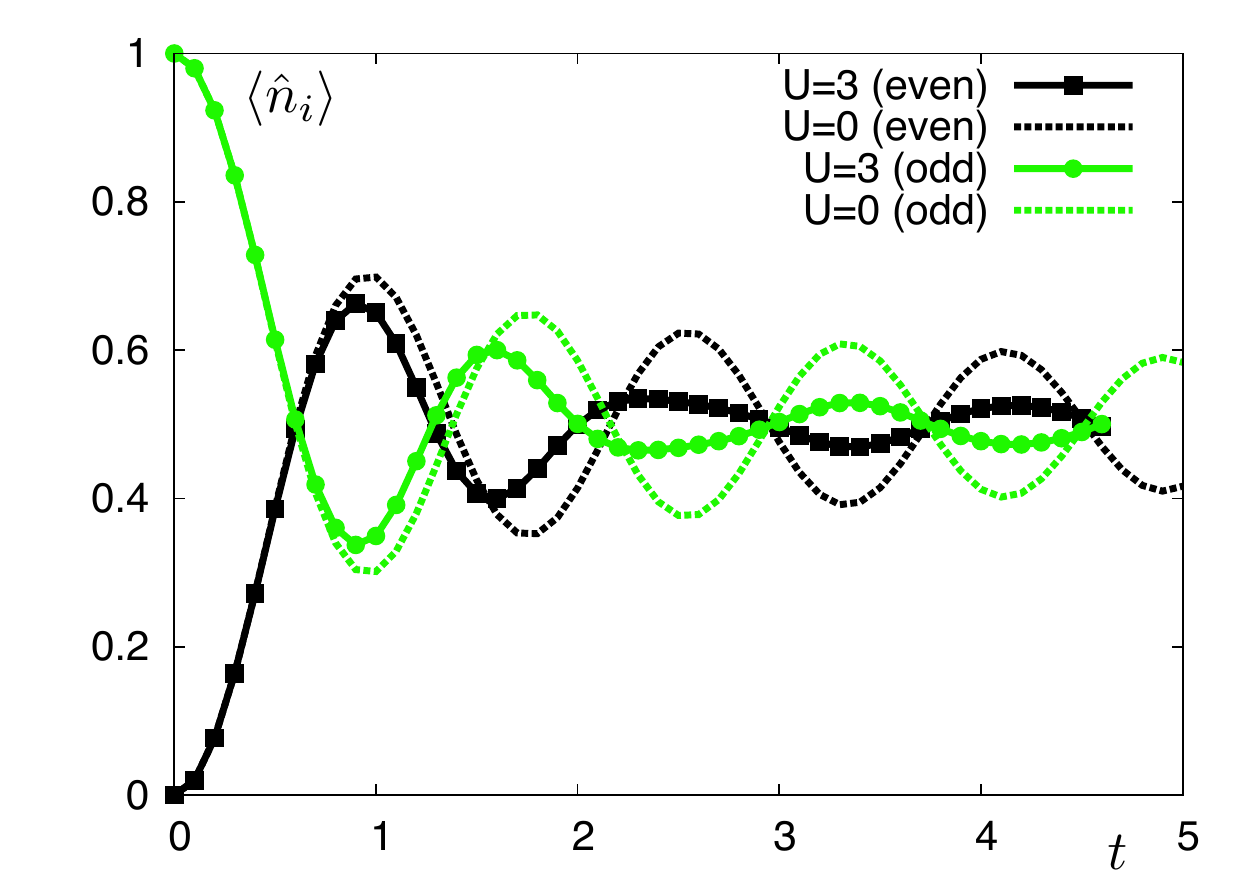}
\caption{\label{F2}Local density $\langle \hat{n}_i (t) \rangle$ vs.\ time, showing local relaxation. Shown is the time evolution of an even and an odd site for $U=0$ and $U=3$. Note the strong deviation from the non-interacting limit and the strong suppression of density oscillations.}
\end{figure}

\begin{figure}
\includegraphics[width=8cm]{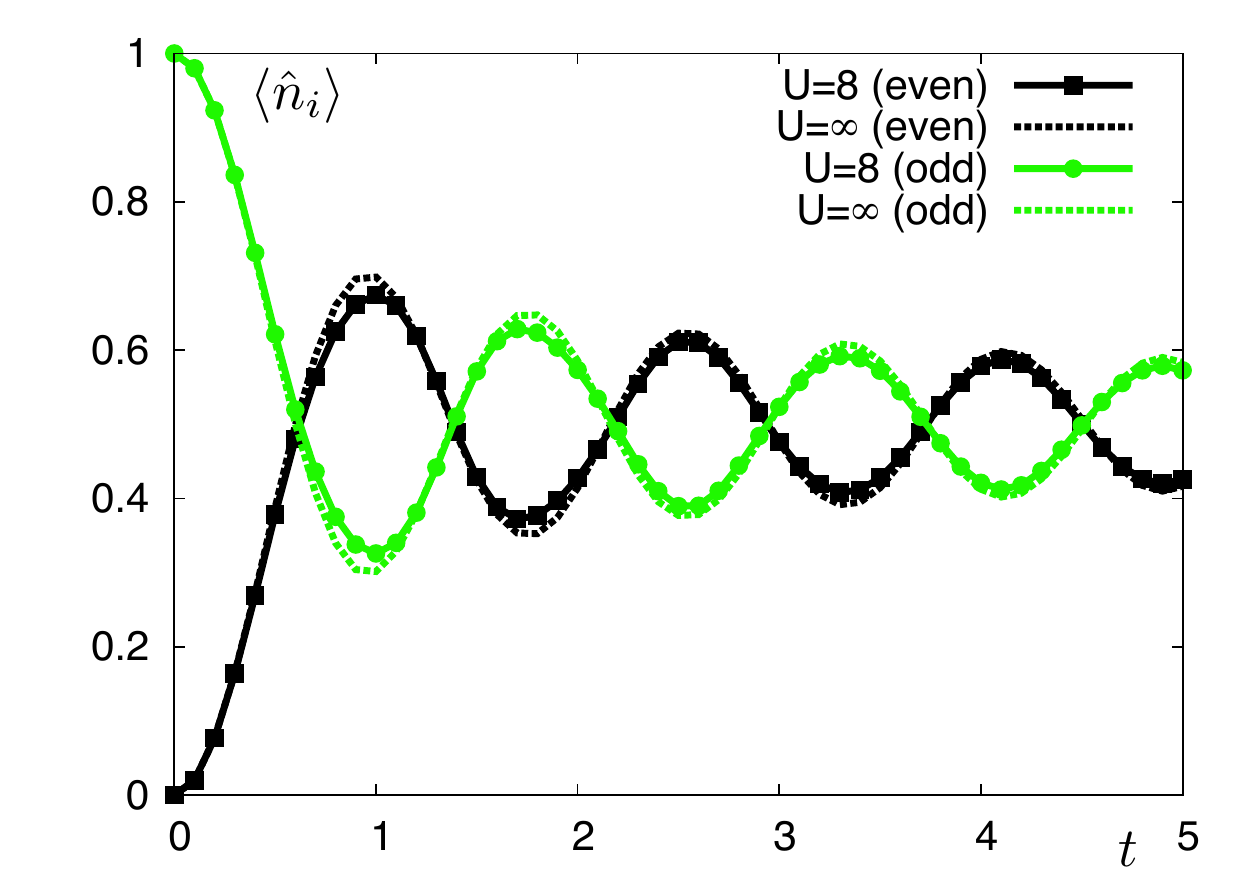}
\caption{\label{F3}Local density $\langle \hat{n}_i (t) \rangle$ vs.\ time, showing local relaxation. Shown is the time evolution of an even and an odd site for $U=8$ and $U=\infty$. The agreement is almost perfect.}
\end{figure}

\begin{figure}
\includegraphics[width=8cm]{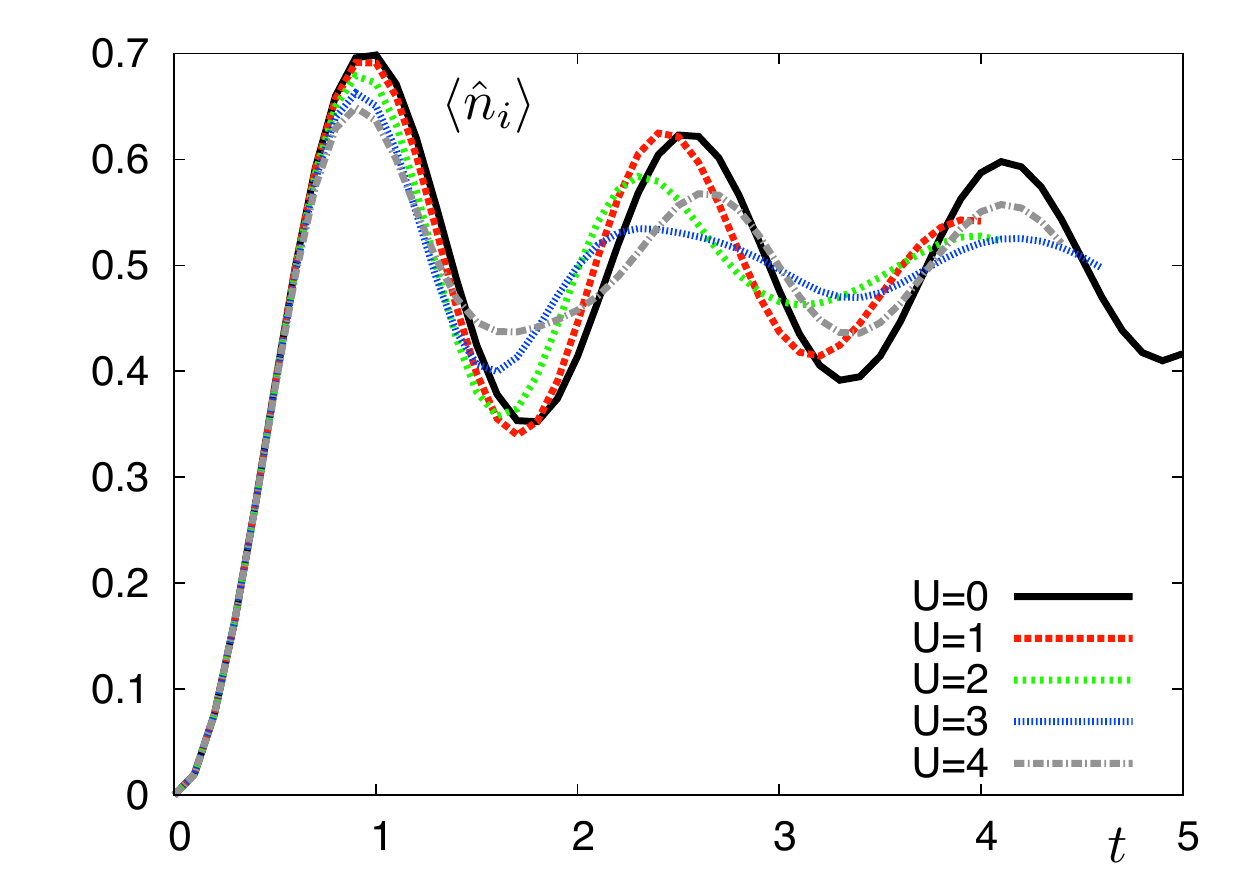}
\caption{\label{realtimedensity1} Time evolution of local densities on even sites  for various $U$ compared to the limiting behavior $U=0$.}
\end{figure}

\begin{figure}
\includegraphics[width=8cm]{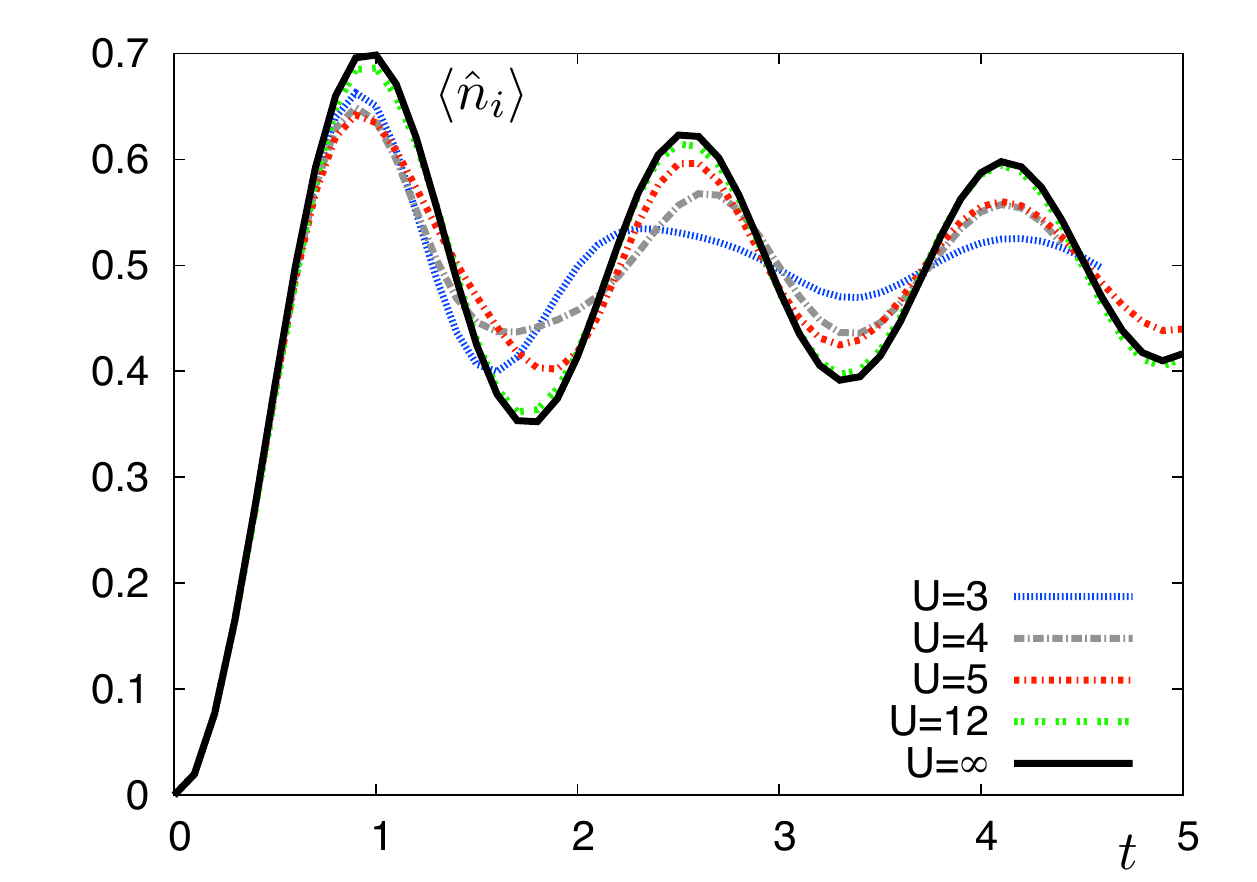}
\caption{\label{realtimedensity2} Time evolution of local densities on even sites for various $U$ compared to the limiting behavior $U=\infty$.}
\end{figure}

All $t$-DMRG results for finite $U$ are perfectly compatible with relaxation of local densities to $n=1/2$.
On very short time scales ($t < 1$) particles have typically
not collided yet and are not yet sensitive to the values 
of $U$, hence, finite $U$ results are identical to the limiting cases $U=0,\infty$.
The relaxation behavior for intermediate times, however, deviates quite strongly.  Interaction effects 
become visible right after particles make contact. 
This can be seen in Figs.~\ref{F1}-\ref{realtimedensity2} (all calculated for $L=32$), where
we compare the time evolution of local densities for various finite values of $U$ to the special cases $U=0, \infty$. For small 
$U$ (exemplified by $U=1.5$) and large $U$ (exemplified by $U=8$), 
the comparison indicates that the relaxation of local densities is 
governed by the behavior of the limiting non-interacting cases. 
For intermediate values of $U$ (exemplified by $U=3$), scattering seems to 
be most effective and lead to much faster damping and relaxation, much beyond
the above square root time dependence. This is plausible,
as close to a quantum 
critical model there is a limiting point in the 
spectrum of the new Hamiltonian, leading to specifically
effective relaxation. Deviations from the limiting 
behavior are sufficiently strong that they  
should be visible experimentally.

\begin{figure}
\includegraphics[width=8cm]{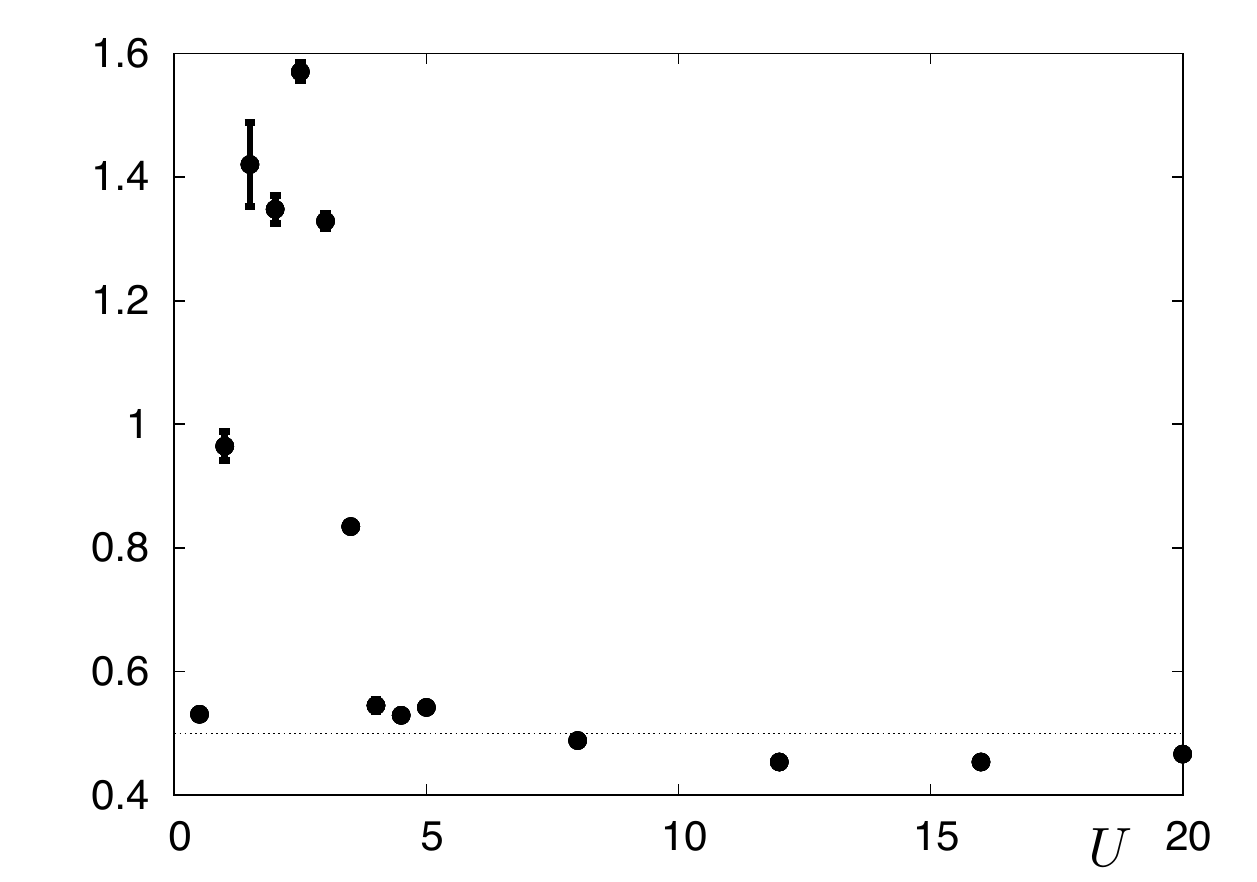}
\caption{\label{slope} Estimated negative
exponents of asymptotic power-law decay law for local densities. 
The asymptotic decay exponent for the limiting cases is shown as a dashed line.}
\end{figure}

Under the assumption that for the time scales achievable by DMRG an asymptotic power-law can already be read off, one 
finds the exponents given in Fig.~\ref{slope}. 
While for very small $U$ and all $U> 4$ one finds a slope 
similar to $t^{-1/2}$, as for the limiting cases, 
there is an intermediate regime where relaxation is much faster. 
It must be stated, however, that in this regime the precise slope is hard 
to extract, even an exponentially fast decay cannot be excluded completely, 
but we are not aware of a physical reason why there should be a qualitative 
change of decay behavior from power-law to exponential. Qualitatively,
stronger interactions should lead to a more distinct dispersion, leading in
turn to a quicker decay.

Concerning experimental implementations, a number of 
further issues require a  
discussion: The experimentally accessible systems 
will be of finite size, leading to recurrence effects. 
Moreover, there will be a parabolic confining potential, which will modify results.
Furthermore, one will encounter initial states at non-zero temperature,
as well as possible additional fluctuations due to 
inhomogeneities in applied fields, different from the
apparent local relaxation.

\begin{figure}
\includegraphics[width=8cm]{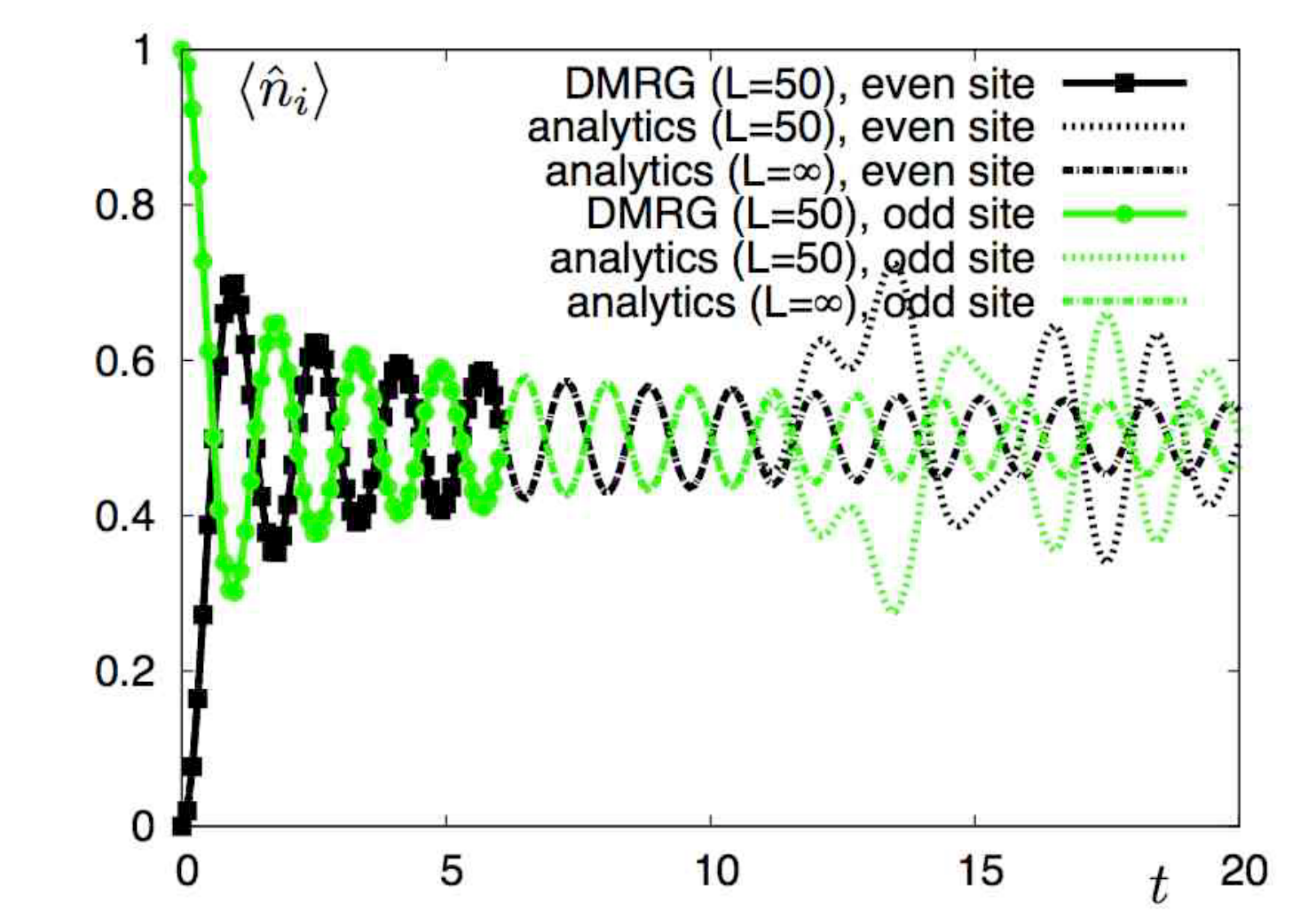}
\caption{\label{recurrencesDMRG} 
Densities in the free limit: Analytics for $L=50$ (PBC), $L=\infty$ (PBC) and DMRG for $L=50$ (OBC; in fact calculated in the hardcore limit, because DMRG fails for $U=0$, but densities are identical in both cases). On the time scales reached by DMRG, all three agree. This incidentally shows that boundary conditions are of little importance for the sizes and time
scales of DMRG. Recurrence effects become visible later.}
\end{figure}

So far, we have confronted analytical results in the $L\rightarrow\infty$ limit with DMRG results for $L=32$. In order to check whether on the time scales reachable by DMRG recurrence effects can be seen for this system size, we have rerun selected calculations for $L=50$, observing no relative change in results above 1 percent. In particular, the shapes of oscillatory behavior remained completely unchanged. In the $U=0$ limit, finite system DMRG results and infinite as well as finite system analytic results agree completely on the finite times reachable by DMRG, even despite the different boundary conditions (open boundary conditions (OBC) for DMRG, 
periodic boundary conditions (PBC) for analytical statements). Recurrences, where the difference between finite and infinite 
system becomes obvious, happen much later, see Fig.~\ref{recurrencesDMRG}.

\begin{figure}
\includegraphics[width=\columnwidth]{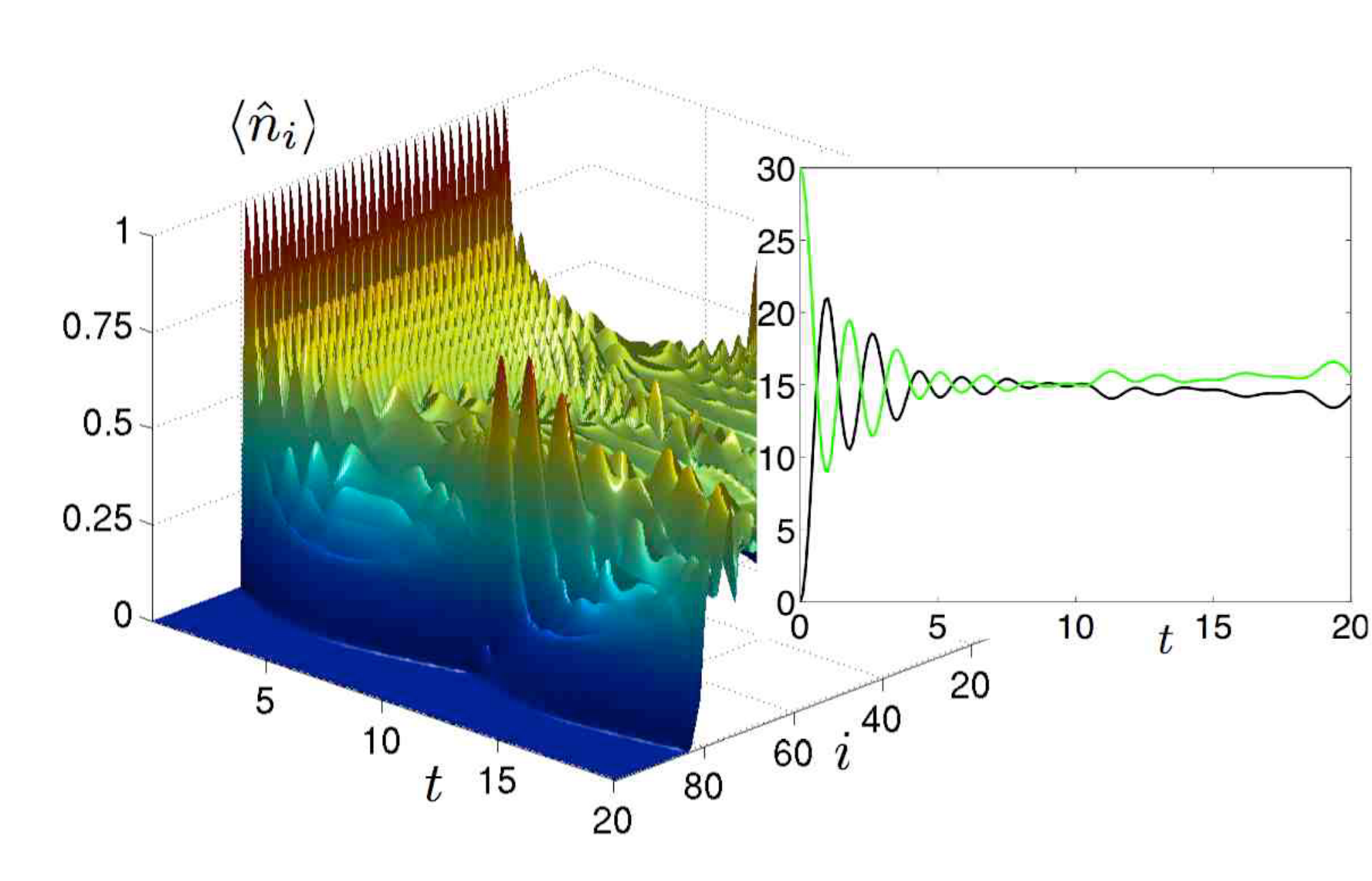}
\caption{\label{density}Density $\langle\hat{n}_i(t)\rangle$ (left) for $U=0$, $L=100$,
including a trapping potential corresponding to a local chemical potential $\mu_i=0.01(i-L/2)^2$. The plot on the right shows $\langle\hat{N}_e(t)\rangle$ (black) 
and $\langle\hat{N}_o(t)
\rangle$ (green). Note the similarity to a Bessel 
function  for short times and the recurrences after 
$t\approx 10$. The recurrences are even more 
pronounced in the quasi-momentum distribution, 
Fig.\ \ref{qmd}, which lacks, however, signatures of 
local relaxation
dynamics.}
\end{figure}

The effect of the trap, as shown in Fig.~\ref{density}, is to generate effective reflections from the edges of the system, leading to much earlier recurrences. However, in experimental setups, this effect would for realistic traps set in late enough to allow for sufficiently long observation.

\begin{figure}
\includegraphics[width=\columnwidth]{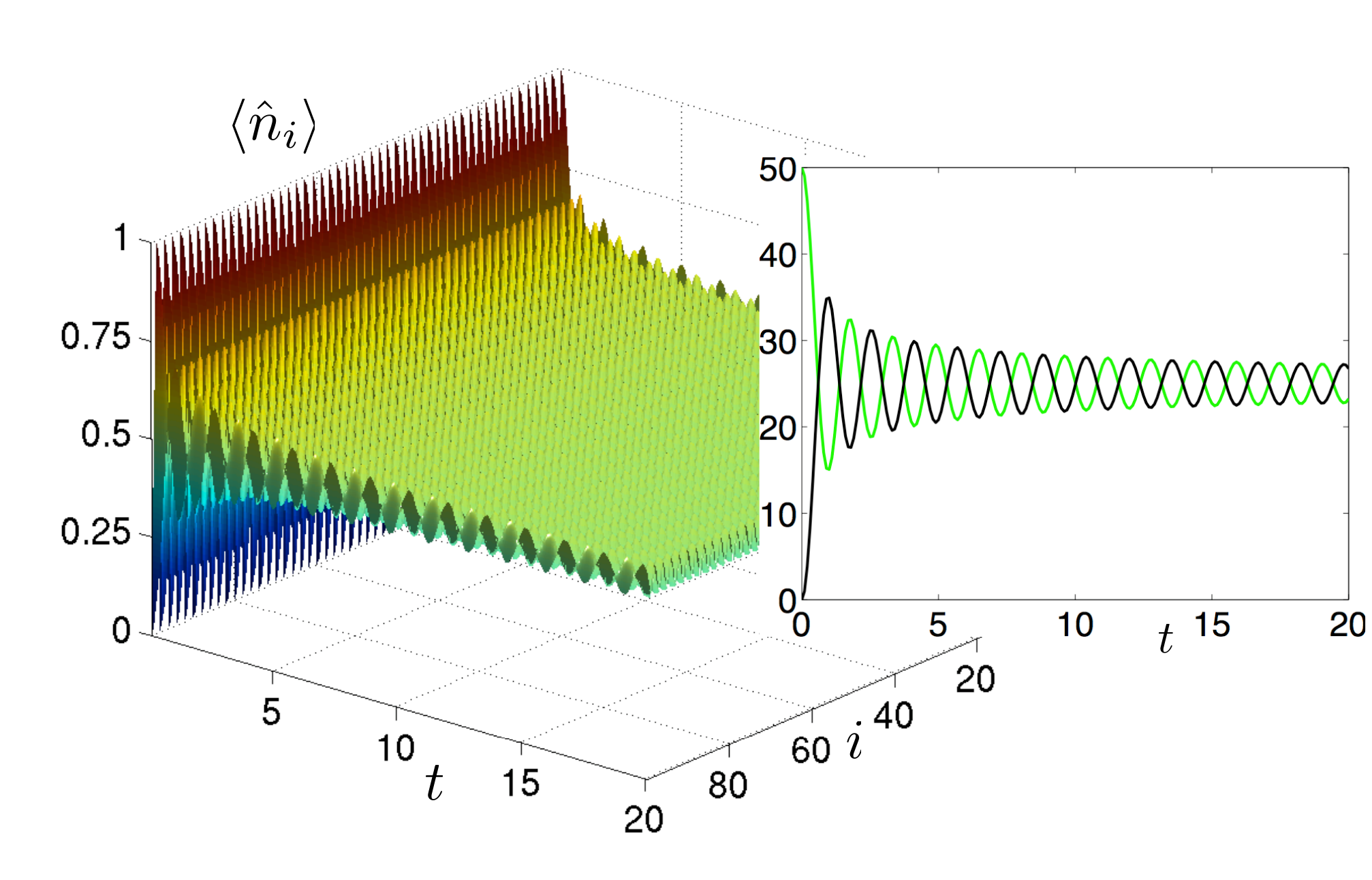}
\caption{\label{densityTi}Density 
$\langle\hat{n}_i(t)\rangle$ (left) for $U=0$, $L=100$, 
and periodic boundary conditions. The plot on the right shows $\langle\hat{N}_e(t) 
\rangle$ (black) and $\langle\hat{N}_o(t)
\rangle$ (green), compare Fig.\ \ref{density}, where 
the same is shown including a trapping potential.}
\end{figure}

\section{Time evolution of local correlators}

Let us now consider the nearest-neighbor correlator $\langle \hat{b}^\dagger_{i+1}(t) \hat{b}_{i} (t)  \rangle$. This quantity is specifically interesting,
and can in principle be measured by means of exploiting the 
tuning of the double well potential of the superlattice, and appropriate
timing (see above, and 
compare also Refs.\ \cite{Wells,TimeResolved,Superlattice}).
It is of interest as it goes beyond local densities: The build-up of correlations in time starting from the uncorrelated initial state becomes visible.

In the limiting free cases, identical results are found:
\begin{equation} 
\begin{split}
\left\langle \hat{b}^\dagger_{i+1}(t) \hat{b}_{i} (t)\right \rangle 
	&=\frac{(-1)^{i}}{2L}\sum_{k=1}^L\me^{4Jt\mi\cos(2\pi k/L)} 
	\me^{-
	 2\pi\mi k /L }\\
	&\rightarrow
	-
	\frac{(-1)^{i}}{2\mi}J_{1} (4Jt) \quad (L\rightarrow\infty).
	\end{split}
\end{equation}
The real part of the correlator is strictly zero for all times, whereas the imaginary part relaxes to $0$ with an asymptotics of $t^{-1/2}$ after a quick growth to a maximal value of about $0.28$ at 
time $t \approx 1/2$. 
This quick growth reflects the buildup of correlations 
due to particle motion with speed linear in $J$. 

\begin{figure}
\includegraphics[width=8cm]{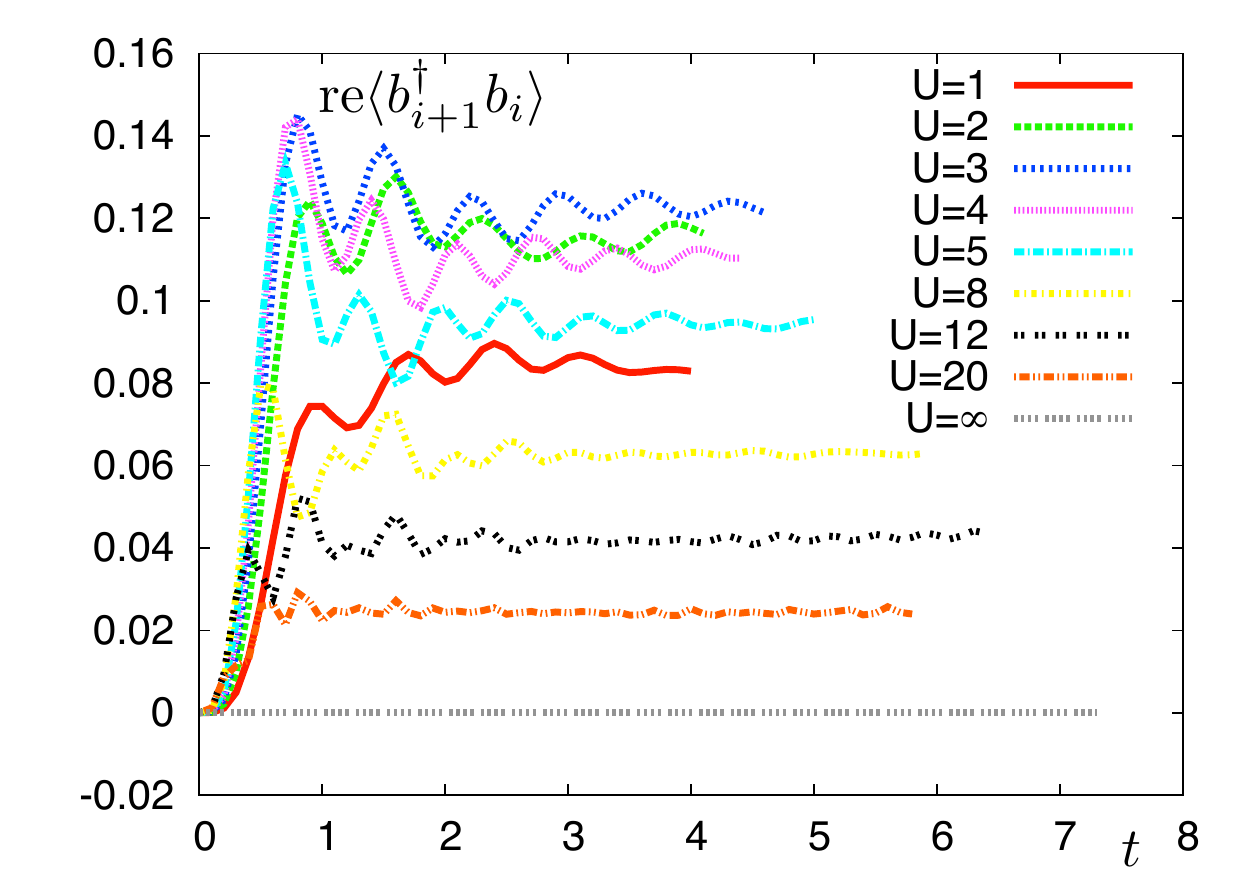}
\caption{\label{realpartnn} Real part of correlations to neighbors 
$\langle \hat b_{i+1}^\dagger (t)
\hat b_i (t) \rangle$ 
as a function of time, for different values of $U$. 
Note that in the effectively free cases $U=0$ and $U=\infty$, 
the absolute value of these correlators converges to zero.}
\end{figure}

\begin{figure}
\includegraphics[width=8cm]{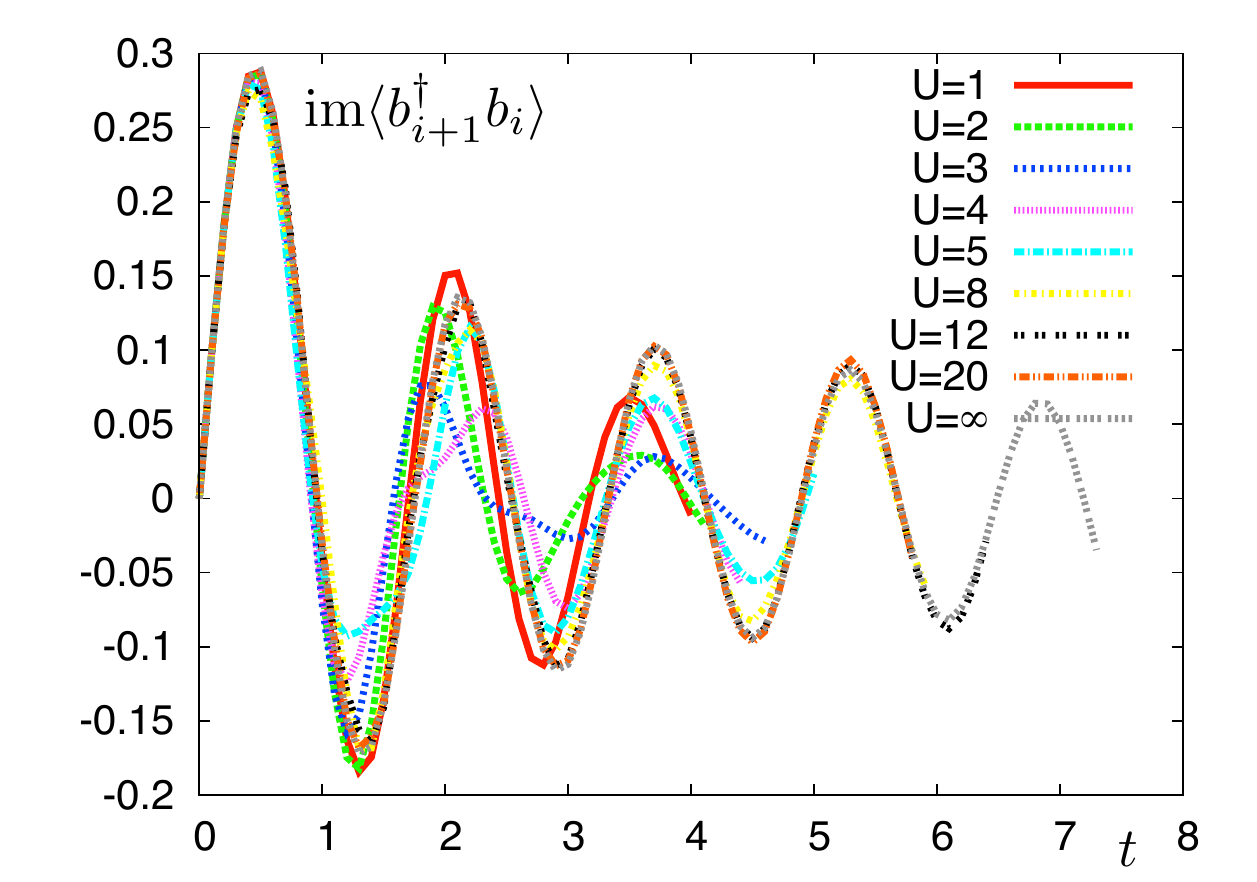}
\caption{\label{imagpartnn} Imaginary part of correlations to neighbors 
$\langle \hat b_{i+1}^\dagger (t)
\hat b_i(t) \rangle$ 
as a function of time, for different values of $U$.}
\end{figure}

For finite $U$, the scenario has marked similarities and differences. Considering Figs.~\ref{realpartnn} and \ref{imagpartnn}, 
one sees that on short time scales the buildup of the imaginary part of correlations is identical for arbitrary $U$. This simply reflects the fact that due to the distance $2$ 
between particles at $t=0$, no collisions have yet happened on these time scales. Only when the interaction strength becomes visible, the relaxation to 0 follows different paths. As for local densities there is a clear trend that relaxation is fastest around $U \approx 3$, reflecting the particularly efficient scattering there. In the real part, convergence to a finite value occurs for all finite $U$, but not such a clear picture of relaxation speeds occurs.

\begin{figure}
\includegraphics[width=8cm]{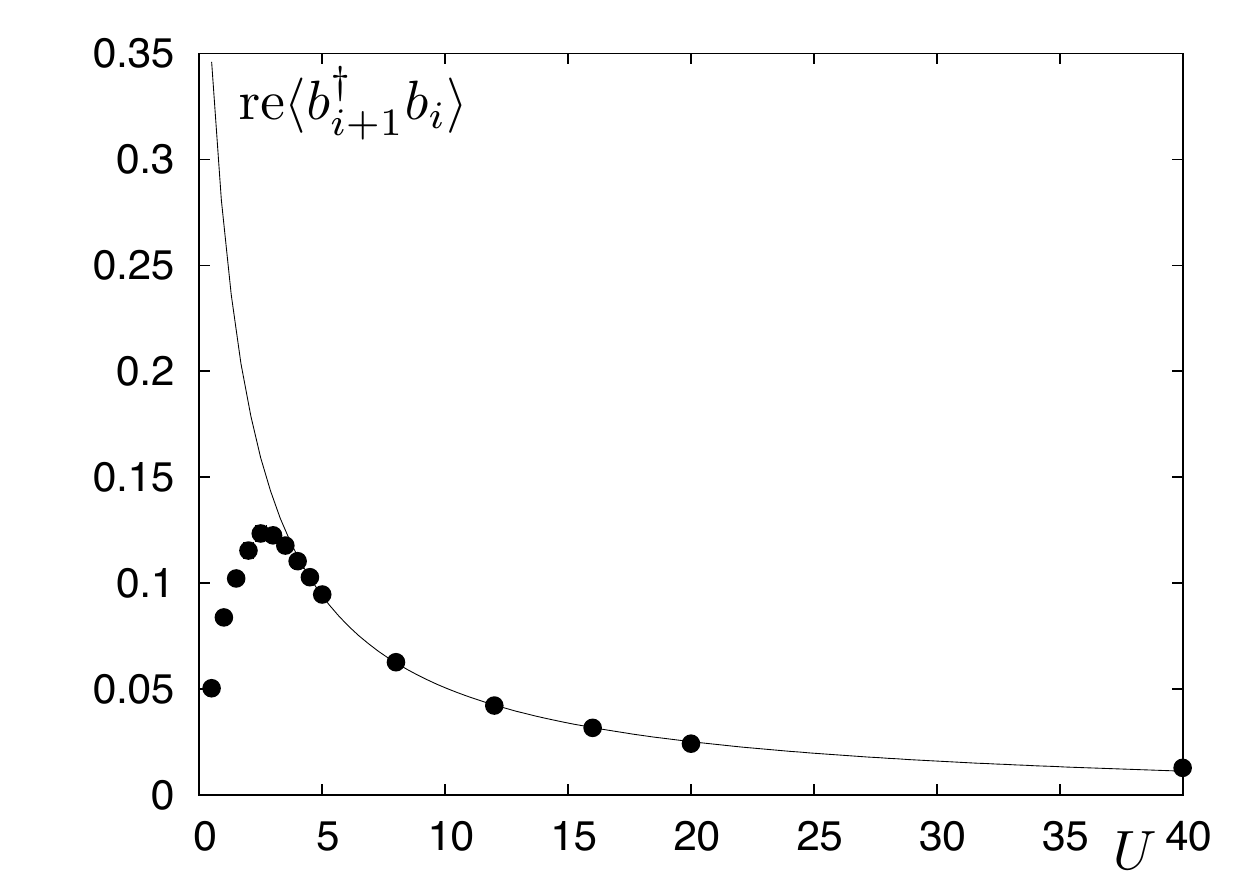}
\caption{\label{realpartnnconv} Equilibrated 
value of the real part of correlations to neighbors 
$\langle \hat b_{i+1}^\dagger (t)
\hat b_i(t)\rangle$ for large $t$ 
for different values of $U$. 
As a guide to the eye, large-$U$ behavior is fitted 
by a solid line proportional to $1/U$.}
\end{figure}

If one plots the asymptotic values of the real part of the 
nearest neighbor correlators (Fig.~\ref{realpartnnconv}), one 
sees that there is a maximum around $U \approx 3$, 
reflecting the efficient scattering there. Interestingly, 
the large-$U$ behavior can be very well approximated by 
a $U^{-1}$ curve for all $U=4$ and larger. 

Indeed, this dependence is exactly what one would expect in the
thermal or Gibbs state of the Bose-Hubbard 
Hamiltonian $\hat H$:
We hence look at local correlations in 
the Gibbs state 
\begin{equation}
	\hat\varrho_\beta=
	\me^{-\beta \hat H}/Z 
\end{equation}
with $Z= \text{tr}[\me^{-\beta \hat H}]$ .
We will take 
\begin{equation}
	\hat{H}_0 = 
	\frac{U}{2}\sum_{i=1}^L
	\hat{n}_i\left(\hat{n}_i-1\right).
\end{equation} 
as our unperturbed Hamiltonian including interactions
and look at leading orders in the hopping term
\begin{equation}
	\hat{V}=-J\sum_{i=1}^L
	(\hat{b}^\dagger_{i+1}\hat{b}_i+ 
	\hat{b}^\dagger_{i}\hat{b}_{i+1}).
\end{equation}
Then we find, using standard thermal perturbation theory, up to first order in $J$
\begin{equation}
\begin{split}
\langle \hat b_i^\dagger \hat b_{j}\rangle& = \delta_{i,j}\langle\hat b_i^\dagger \hat b_{i}\rangle
-\int_{0}^\beta\!\!\!\!\md x\,\frac{\text{tr}[\me^{(x-\beta)\hat{H}_0}\hat{V}\me^{-x\hat{H}_0}\hat b_i^\dagger \hat b_{j}]}{Z}.
\end{split}
\end{equation}
This means that we have ($\delta_{\langle i,j\rangle}=1$ if $i$ and $j$ are nearest neighbors, zero otherwise)
\begin{eqnarray}
	\langle \hat b_i^\dagger \hat b_{j}\rangle = \delta_{i,j}\langle\hat b_i^\dagger \hat b_{i}\rangle
	\hspace{5.5cm}
	\\
	+\delta_{\langle i,j\rangle}
	\frac{J}{U}\sum_{n,m}\me^{-\beta(E_n+E_m )}n(m+1)\frac{\me^{\beta U(n-m-1)}-1}{z^2(n-m-1)}
	\nonumber
\end{eqnarray}
where $z= \sum_{n} \me^{-\beta E_n}$ and $E_n=Un(n-1)/2$ are the 
local energies of the unperturbed Hamiltonian.
So, indeed, within the validity of
perturbation theory, we do find the anticipated 
linear dependence on
$1/U$, as seen also in DMRG simulations in the time-dependent
scenario. By definition, the 
correlators merely probe local quantities.
This corroborates the intuition that locally, 
the system is indistinguishable from the situation as if 
globally the system was in a state maximizing the entropy, 
under the constraints of motion \cite{Relax,Tegmark,Barthel}. 

\begin{figure}
\includegraphics[width=8cm]{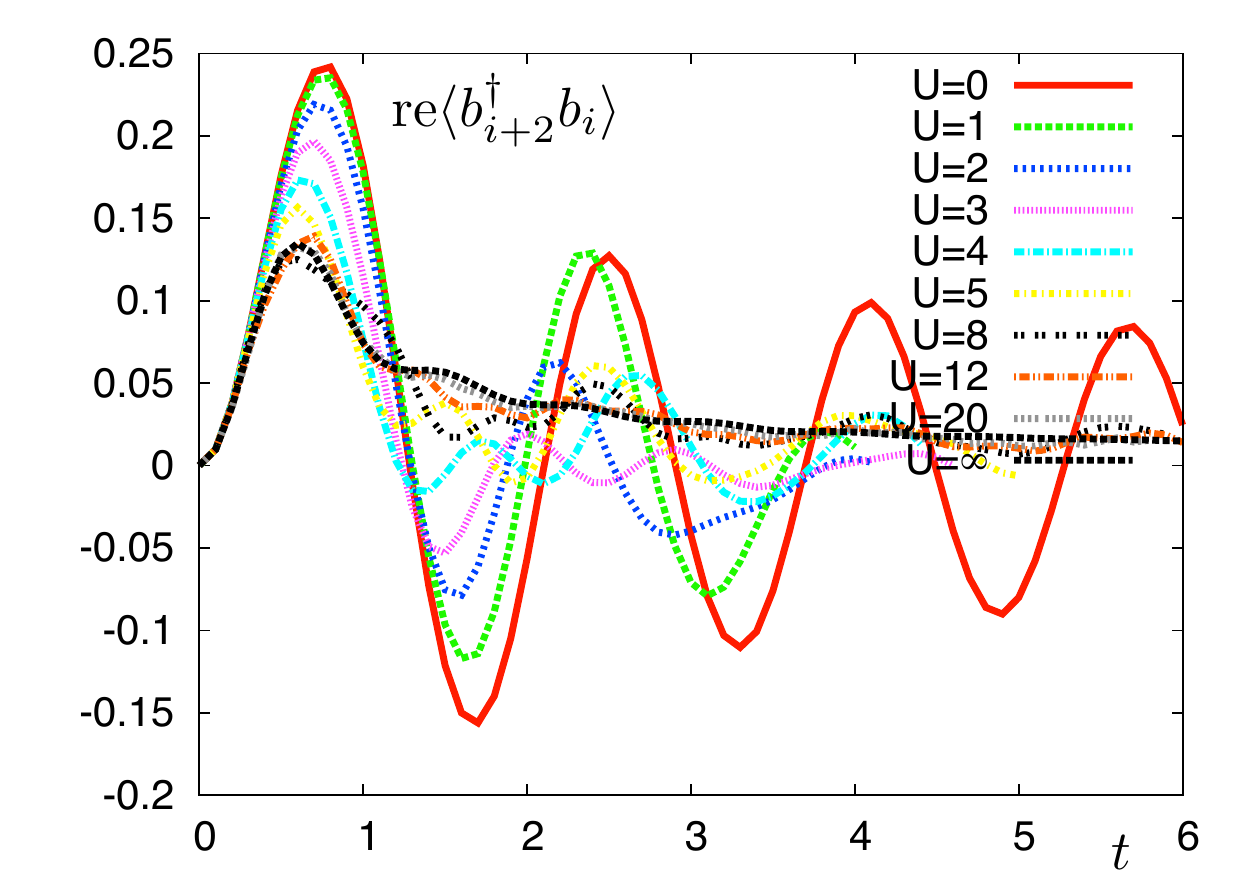}
\caption{\label{realpartnn2} Real part of correlations to neighbors 
$\langle  \hat b_{i+2}^\dagger (t) \hat b_i (t) \rangle$ 
as a function of time, for different values of $U$.}
\end{figure}

To complete the picture, let us finally remind ourselves of how
the maximum entropy states of the global system 
locally look like. 
The constants of motion of interest here 
are the occupation numbers in bosonic or 
fermionic momentum space, respectively. 
The global maximum entropy state---under the constraints of motion---for 
$U=0$ is then found to be
\begin{equation}
	\hat{\varrho} = \bigotimes_{i=1}^L 
	\frac{2}{3} e^{-\ln(3)\hat{b}^\dagger_i \hat{b}_i}.
\end{equation}
This result is completely consistent 
with our result for the relaxation of the characteristic function (see Appendix \ref{sec:proof_relax}).  
One can now calculate expectation values in equilibrium for the \emph{local} observables considered above (by construction in agreement 
with our earlier analytical findings) as
\begin{equation}
	\langle \hat{b}^\dagger_i \hat{b}_j \rangle  = 
	 \frac{1}{2} \delta_{i,j}	,\,
	\langle \hat{n}_i \hat{n}_j \rangle - \langle \hat{n}_i 
	\rangle \langle \hat{n}_j \rangle =\frac{3}{4}\delta_{i,j}.
\end{equation}
In the case of $U \rightarrow \infty$ one finds
\begin{equation}
	\hat{\varrho} = \frac{1}{2^L} \id,
\end{equation}
which means that
\begin{equation}
	\langle \hat{b}^\dagger_i \hat{b}_j \rangle 
	= \frac{1}{2} \delta_{i,j},\,\,
	\langle \hat{n}_i \hat{n}_j \rangle - 
	\langle \hat{n}_i \rangle \langle \hat{n}_j \rangle =
	\frac{1}{4}	\delta_{i,j}.
\end{equation}
We have also numerically investigated 
the relaxation behavior of longer-ranged 
correlators $\langle  \hat b_{j}^\dagger (t)
\hat b_i (t) \rangle$. As an example, 
we depict in Fig.~\ref{realpartnn2}
the relaxation of the real part of 
$\langle \hat b_{i+2}^\dagger (t)
\hat b_i (t) \rangle$, the imaginary part being zero.
Relaxation here is of monotonically increasing 
effectiveness with $U$. The relaxation dynamics 
for $U=0$ and $U=\infty$ is different, revealing the 
fundamentally different character of the non-interacting 
limiting cases.

\section{Local relaxation}

Let us summarize the properties of local relaxation that can be extracted in this experimental setup. In all quantities, there are three {\em time regimes:}

\begin{enumerate}
\item The first time regime is associated with the {\em building up of 
correlations}: This is governed directly by the coupling
strength to the nearest neighbor, and on the shortest times is 
identical for all $U$, before collision processes become important. The time scale is $Jt<1$, as this is when the first collisions happen and establish correlations.

\item The second time regime is associated with {\em local relaxation}. There is fast oscillating dynamics between neighboring sites, happening at a time scale dictated by
the hopping $J$, which also governs the speed of sound. 
Local relaxation as such is slow, in the exact analysis a slow polynomial decay:
The Bessel function fulfills $|J_0(x)|\leq x^{-1/3}$ as a strict bound, with an asymptotic envelope 
\begin{equation}	
	J_0(x) = x^{-1/2}+ o(x^{-1/2}).
\end{equation}
one finds relaxation ({\em not} due to decoherence but due the 
dilution of information over the lattice) to the true local 
equilibrium. For finite $U$, numerics is consistent with 
polynomial decay, which for intermediate $U$ 
seems to be much {\it faster}, but still polynomial.

The intuition is that this local relaxation is due to influences
of excitations travelling with the speed of information 
propagation from farther
and farther lattice sites, broadened by {\it dispersion}. 
This means that this time scale,
even in the interacting case of $U>0$, should be
defined by the {\it speed of information propagation} in the system. 
Note that to date, there is no rigorous bound known to 
the speed of information propagation in the fully interacting Bose-Hubbard
model (this being a consequence of the lattice sites having
an unbounded number of particles, see also comments in
Refs.\ \cite{Nachtergaele,HarmonicLiebRobinson,Buer}).
On physical grounds, yet, it should be expected to be 
similar to $J$, i.e.\ "ballistic" transport at the high energies provided by the quench, slightly modified by $U$ (possibly the 
resulting bounds can only be formulated for specific 
initial states having small local particle number). 
This is also what the numerics shows. Note that the low-energy speed of sound need not be relevant here.

\item The third (very large) time is the {\it recurrence time}, which seems to be shortened substantially in the presence of a realistic trap. As in the presence of
a trap, the excitations no longer travel with a constant 
speed (the speed of sound), but are slowly reflected, one
should expect a quicker relaxation, and this is also the
behavior the analytics exhibits. 
Generally, for moderate to large system sizes it is already 
beyond the reach of our simulations. 
It would be interesting to see,
possibly in exact diagonalization, whether the recurrences are
weakened compared to the free solutions
due to interactions in the system, see also Ref.\
\cite{LightCone}.

\end{enumerate}

While these findings are essentially independent of the interaction strength $U$, we also find three local relaxation regimes for different $U$: 
\begin{itemize}
\item For very small values of $U$ (up to $U\approx 1$) 
relaxation dynamics is quite close to the non-interacting bosonic limit.

\item For larger values of $U$ the system seems to show a behavior very similar to the hard core bosonic or free fermionic limit with similar relaxation exponents of order $1/2$. 
Local correlations are consistent with locally 
equilibrated subsystems perturbatively coupled in $1/U$.

\item The case of intermediate $U \sim 3$ appears to be a special case for various observables, 
marking the
``boundary'' between the ``free bosonic case'' 
$U=0$ and the 
``hard core
boson case'' $U=\infty$, so the fermionic one. 
Collisions lead to the most efficient relaxation in this case. 
\end{itemize}
To summarize, the dynamics of local quantities shown is consistent with the limiting  
$U=0$ and $U=\infty$ cases, but shows a richer phenomenology in particular for intermediate $U$.

\section{Quasi-momentum distribution}

In this subsection, we briefly compare what is seen in the above
local relaxation with the situation in momentum space. The quasi-
momentum distribution,
\begin{equation}\label{qm}
	S(q,t)=\frac{1}{L}
	\sum_{i,j=1}^L \me^{\mi q(i-j)}\langle \hat b_i^\dagger (t)
	\hat b_j (t)
	\rangle,
\end{equation}
obtainable from time-of-flight experiments,
is no 
longer a local quantity, but a global one probing the state of
the entire lattice, as well as the boundary conditions and possibly
a confining additional potential.
In fact, in the free model for $U=0$, one finds that the 
quasi-momentum distribution
will be very little time-dependent, and if so, in a chaotic 
fashion showing little structure, see Fig.\ \ref{qmdTi}.
More specifically, we find (see Appendix) $S(q,t)=1/2$ for $q=2\pi l/L$, $l=1,\ldots, L$,
and
\begin{equation}
	S(q,t)=\frac{1}{2}+\frac{\mi}{L^2}\sum_{p=1}^L
	\me^{4J\mi t\cos(2\pi p/L)}\frac{\sin^2(Lq/2)}{\sin(2\pi p/L-q)}.
\end{equation}
for all other $q\in[0,2\pi]$. 
In case of $U=0$, the quasi-momentum distribution hence indeed
does not show any signatures of the local dynamics, see Fig.\ \ref{qmdTi}. 
This is, again, not inconsistent at all with the notion of local relaxation: 
This quantity is a global one (and local only insofar as correlators
of far away lattice sites are suppressed via a quickly rotating
exponential function). 
For any subblock, the dynamics drives the  
system towards the values of equilibrium. In momentum space,
this is not necessarily 
the case; at least not on these time scales. In the free situation of $U=0$, 
this leads to an absence
of relaxation of this quantity. 

It would still be interesting to measure this quantity for close
to free cases for short times, to demonstrate the 
very point that the information about the initial condition 
is fully contained in the system, and merely diluted. Hence, obviously, 
one cannot expect true global relaxation
to happen, unless a further mechanism of 
decoherence due to additional external degrees of freedom 
is present. 

\begin{figure}
\includegraphics[width=\columnwidth]{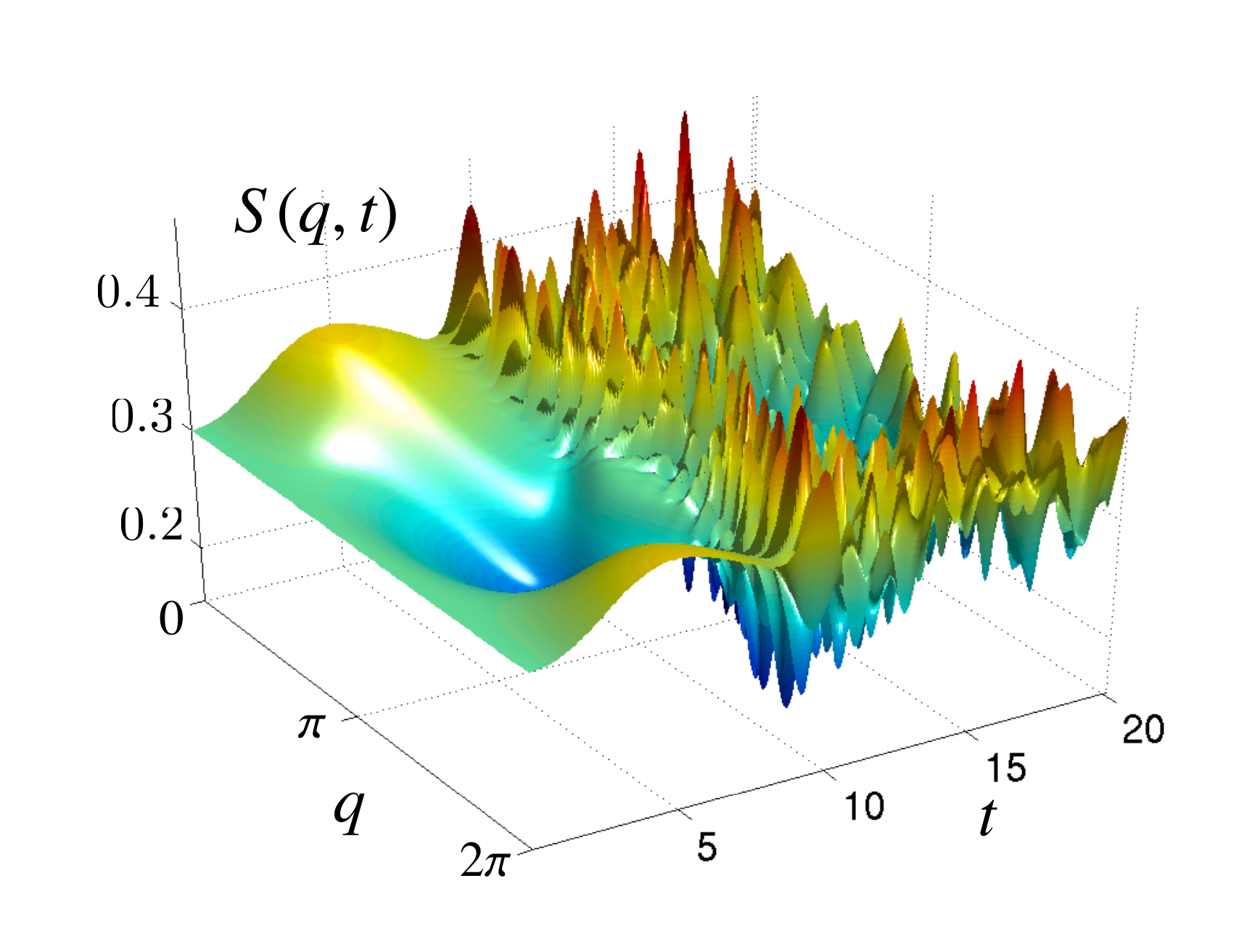}
\caption{\label{qmd}Quasi-momentum distribution $S(q,t)$ for the same parameters as in Fig.\ \ref{density}. Note the absence of the time-scales clearly visible for local observables as in Figs.\ \ref{densityTi}, \ref{density}: The fast oscillatory behavior and the relaxation. The recurrence is however clearly visible.}
\end{figure}

In Fig.\ \ref{qmd} we show
the quasi-momentum distribution 
for the situation including a trapping 
potential. The effect of the additional harmonic potential is 
very significant. Yet, again, note the absence of signatures
of the time  scales of local dynamics as compared to 
Figs.\ \ref{densityTi}, \ref{density}.

\begin{figure}
\includegraphics[width=8cm]{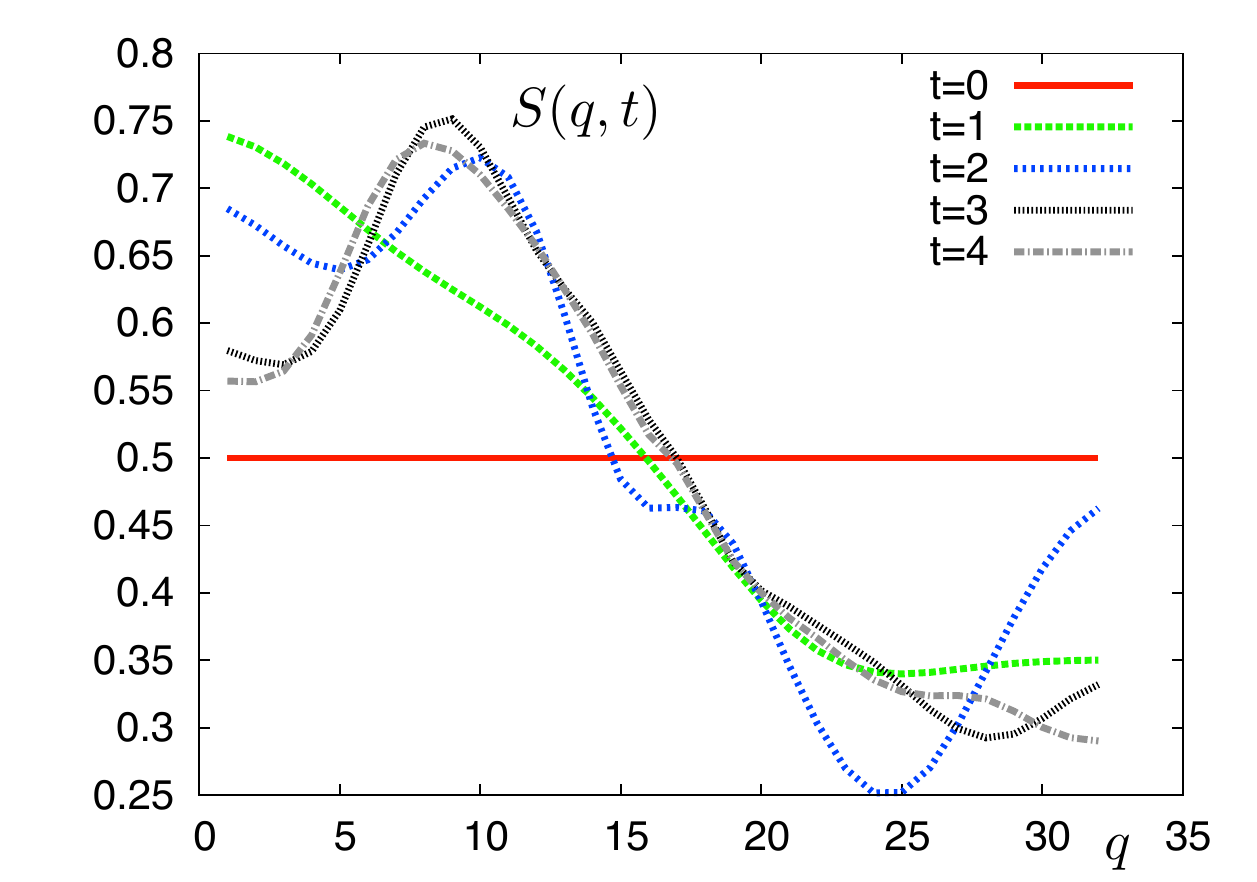}
\caption{Quasi-momentum distribution as a function of time, for
$U=1.5$. The labels $q$ on the $x$-axis correspond to momenta 
${\pi}/({L+1})q$, 
best suited to DMRG open boundary conditions, as explained in the text.}\label{modi}
\end{figure}

\begin{figure}
\includegraphics[width=8cm]{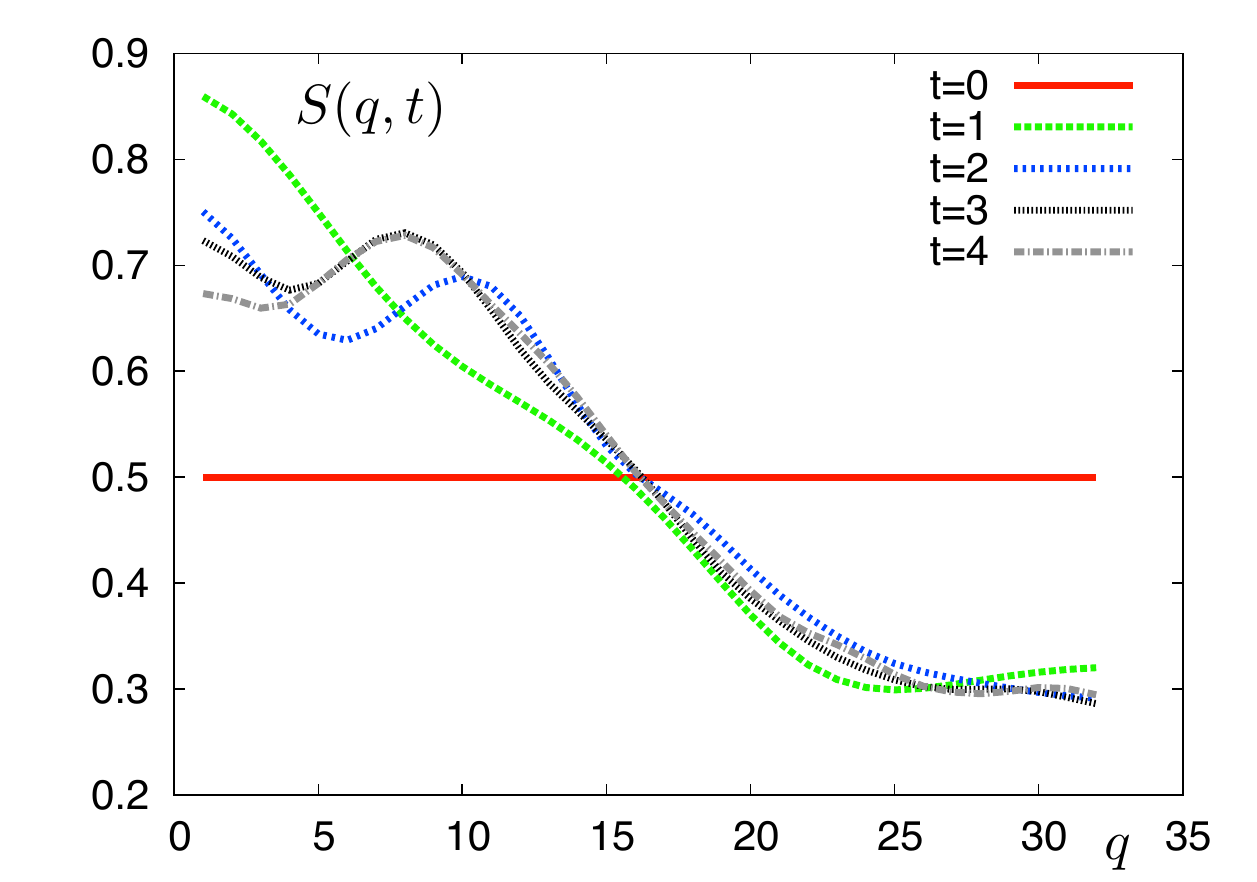}
\caption{Quasi-momentum distribution as a function of time, for
$U=3$, marking the cross-over regime. Labels as in Fig.~\ref{modi}.}\label{modi2}
\end{figure}

How does this compare to the case of having a finite on-site interaction $U$?
Again, the behavior reminds of the situation in the 
free case of $U=0$. The quantity is too non-local to grasp the
effect of local relaxation.
Still, for longer times, one may also expect
relaxation in some translationally invariant properties
in momentum space, and the quasi-momentum distribution 
may well relax. Such a situation is observed 
in a flow equations approach in case of an interacting model in 
Ref.\ \cite{Moeckel}. For the DMRG calculations we use a slightly different definition of the quasi-momentum distribution
\begin{equation}
S(q,t) = \frac{2}{L+1} \sum_{i,j=1}^L \sin\left[\frac{\pi qi}{L+1}\right] \sin\left[\frac{\pi qj}{L+1}\right] \langle \hat{b}^\dagger_i(t) \hat{b}_j(t) \rangle.
\end{equation}
This ensures that $S(q,t)$ is a constant of motion for $q=1,\dots,L$ in the free system $(U=0)$ in case of OBC. Fig. \ref{modi} and Fig. \ref{modi2} show this quasi-momentum distribution for $q=1,\dots,L$ for different values of $U$. Indeed, 
signatures of such a behavior are also observed in Fig.\ \ref{modi},
where the momentum distribution is depicted for different values of
time $t$. For larger times, the  quasi-momentum distribution appears to 
relax in the interacting case of $U=1.5$. This effect is even more
distinct in the case of stronger interactions, as depicted in Fig.\ \ref{modi2}.

To reiterate, the very absence of relaxation in close to free
settings on short times gives rise to an interesting situation: One  could 
experimentally see signatures of local relaxation. One could,
however, also see that the information becomes more and
more dilute in position space, but is still preserved in the
system as such, as seen in the quasi-momentum distribution.

\begin{figure}
\includegraphics[width=\columnwidth]{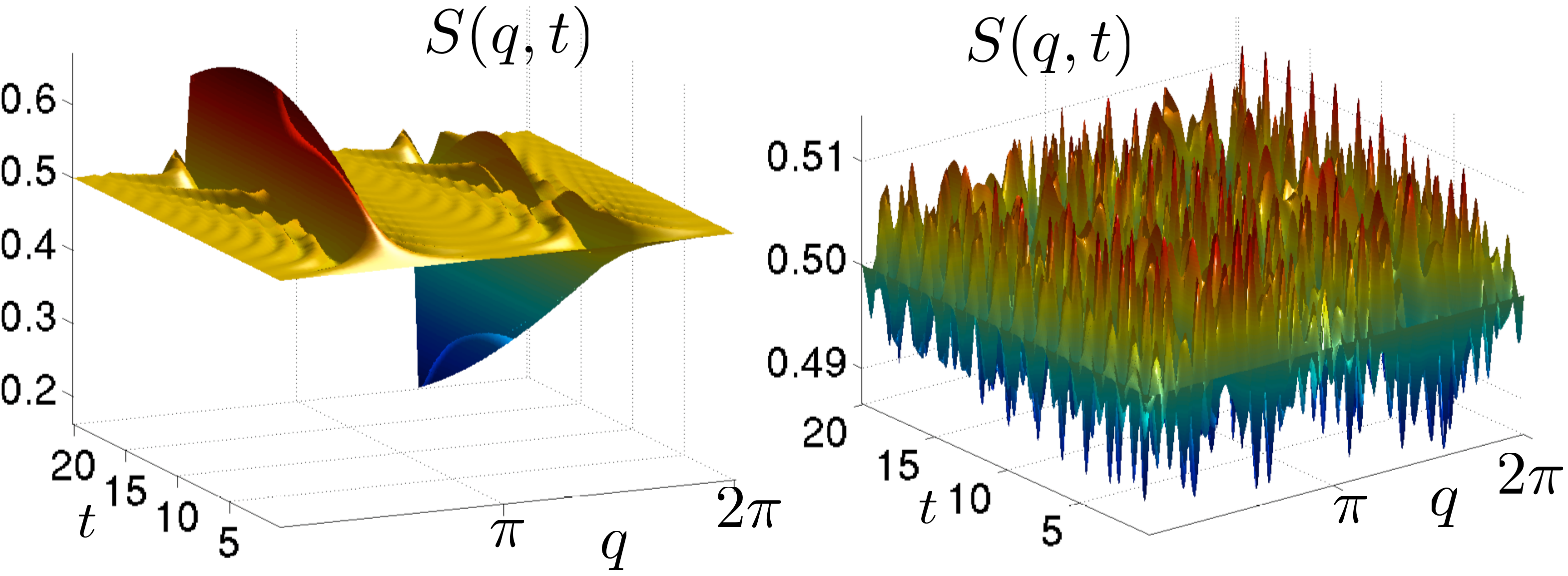}
\caption{\label{qmdTi}Quasi-momentum distribution $S(q,t)$ for 
$U=0$, $L=100$, and open (left), periodic (right) boundary conditions.
Note the dramatic difference between open/periodic boundary 
conditions and the absence of the time-scales clearly visible for local 
observables as in Figs.\ \ref{densityTi}, \ref{density}: The fast 
oscillatory behavior and the relaxation time. See Fig.\ \ref{qmd} 
for $S(q,t)$ in the presence of a trap. The non-local quantity of the
quasi-momentum distribution detects the boundary
conditions early.}
\end{figure}

\section{Global density-density correlator}

\begin{figure}
\includegraphics[width=8cm]{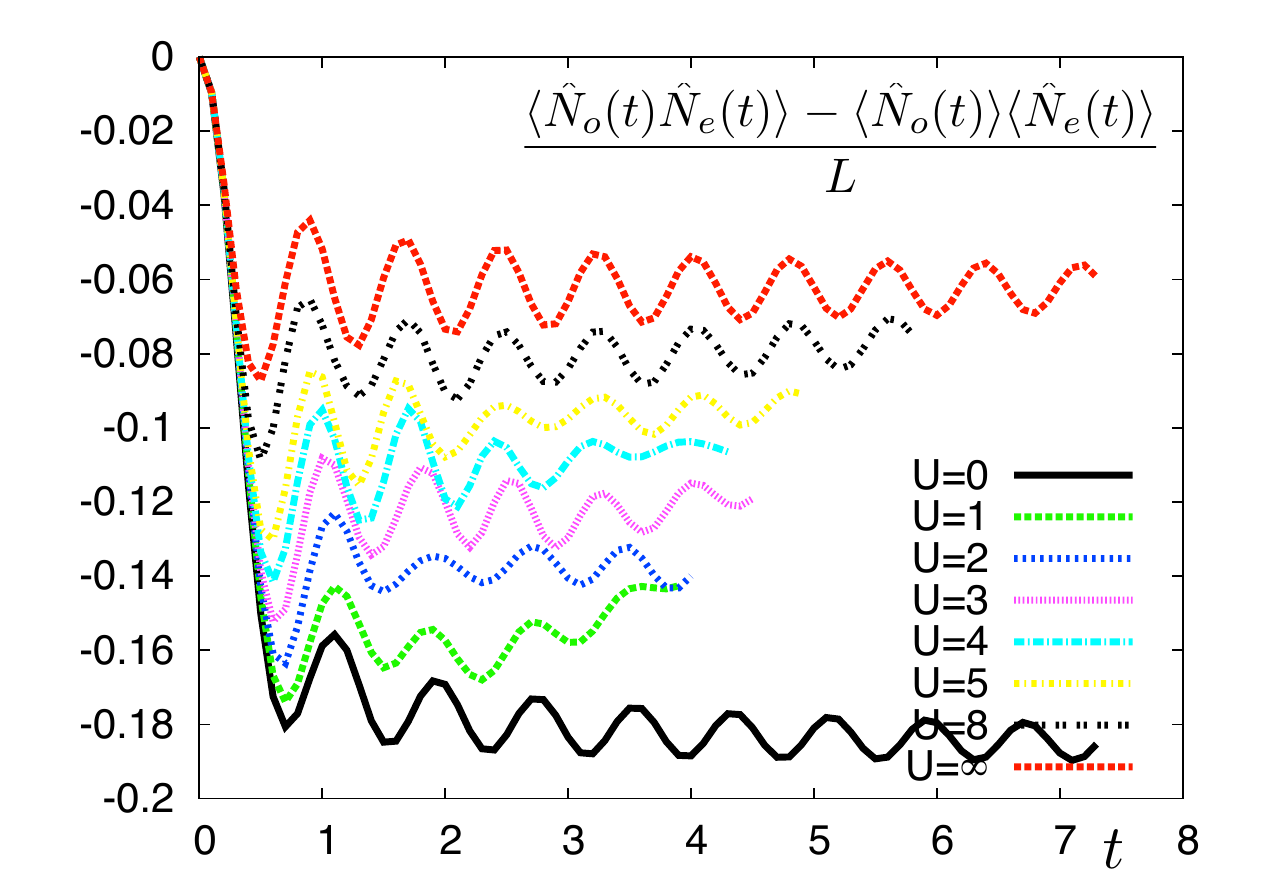}
\caption{\label{dendencorrelations}Global density-density correlations 
for a system of length $L=32$ for various interaction strengths $U$.}
\end{figure}

One potentially experimentally accessible  global correlator is the connected density-density correlator
\begin{equation}
	\langle\hat{N}_e(t)\hat{N}_o(t)\rangle-\langle\hat{N}_e(t)\rangle\langle\hat{N}_o(t)\rangle 
\end{equation}
which we show for different $U$ in Fig.~\ref{dendencorrelations}. 
Numerics indicates that this quantity looks effectively relaxed after some 
time as long-range correlators contributing will be small. For the free
case of $U=0$, one analytically finds for large $L$ (see Appendix)
\begin{equation}
\begin{split}
&\langle\hat{N}_e(t)\hat{N}_o(t)\rangle-\langle\hat{N}_e(t)\rangle\langle\hat{N}_o(t)\rangle\\
&\hspace{0.4cm}\rightarrow
-\frac{L}{16}\left(
3+
J_0(8Jt)
-4[J_0(4Jt)]^2\right),
\end{split}
\end{equation}
which relaxes to $-3L/16$ for large times. For hardcore bosons the global density-density correlator for large $L$ is given by
\begin{equation}
\begin{split}
&\langle\hat{N}_e(t)\hat{N}_o(t)\rangle-\langle\hat{N}_e(t)\rangle\langle\hat{N}_o(t)\rangle\\
&\rightarrow - \frac{L}{16} \left(1-J_0(8Jt)\right),
\end{split}
\end{equation}
which relaxes to $-L/16$ for large times.

\section{Entanglement and entropy growth}\label{ED}

The creation of excitations at all points of the lattice
will in general create a significant degree of entanglement in 
the time evolving state. Since at each site, an excitation
starts to travel through the lattice, a linear increase of the 
entanglement entropy of subblocks in time is to be 
expected \cite{Entanglement0,Entanglement,Entanglement2}.
More precisely, it has been shown in 
Refs.\ \cite{Entanglement,Entanglement2} 
that for any time evolution under a local Hamiltonian with finite-dimensional
constituents, starting from a product state, the entanglement entropy of a subblock
$I= \{1,\dots, s\}$ grows in one-dimensional systems in $s$
at most as
\begin{equation}
	E(\hat\varrho_I(t)) \leq c_0 t + c_1,
\end{equation}
for suitable constants $c_0, c_1$. This means that for any constant time,
the entanglement entropy satisfies what is called an {\it area law}.
For larger and larger block sizes, the entanglement entropy will
eventually saturate in $s$ (and, e.g., not logarithmically diverge).

In turn, this {\it upper bound} is
saturated, in the following sense: In  
Ref.\ \cite{Relax} and
explicitly in the Appendix 
$1$-norm convergence is shown to 
products of Gaussian states 
\begin{equation}
	\| \hat\varrho_I(t) - \hat\varrho_{\text{max}}\|_1=
	\| \hat\varrho_I(t) - \hat\varrho_G^{\otimes s}\|_1\rightarrow 0,
\end{equation}
when considering initial conditions $|\psi(0)\rangle\langle(0)|$ and
time evolution under $\hat H_{U=0}$, where $\hat\varrho_{\text{max}}$ is a 
Gaussian product state. This observation
immediately implies a bound to the entanglement entropy that is linear in time,
if the block size $s$ can only 
be chosen appropriately. This has also been 
made explicit in Ref.\ \cite{Schuch}, in that for the 
fermionic instance of this problem (or, equivalently, a spin model),
there exist constants $c_2,c_3,c_4, L_0, s_0, t_0 >0$ such that 
\begin{equation}
	E(\hat\varrho_I(t)) \geq c_2 t + c_3,
\end{equation}
for $L\geq L_0$ and $s\geq s_0$ and $t_0\leq t\leq c_4 s$,
again for $I=\{1,\dots, s\}$.
In other words, for a larger and larger block size, one encounters
a linear increase of the entanglement entropy of that block. 
In both Ref.\ \cite{Relax} and \cite{Schuch} the local states converge 
to maximal entropy states under the constraints of motion, in 
the latter case starting from a fermionic Gaussian state.

This means that eventually, one will have to use matrix-product
states in the DMRG approach that make use of exponentially
many parameters in time. This linear increase, provably correct
in the above cases, is consistent with a wealth of numerical
findings in quenched systems, as well as with arguments using
conformal field theory \cite{Daley,Entanglement3,Entanglement4,Calabrese}. 

Knowing that we have to asymptotically deal with a linear increase,
we have plotted the entanglement entropy as obtained from a
t-DMRG approach. We depict here the entanglement entropy 
as a function of time in a slightly different geometrical setting, 
which still has implications on the approximatability of a 
state with a matrix-product state in DMRG. This is the setting
of the symmetrically bisected half-chain, for $s=L/2$ and  $L=32$. 
The above linear lower bound is also rigorously true for 
this geometry (at least for finite-dimensional constituents),
whereas the linear upper bound is certainly expected to be valid.

\begin{figure}
\includegraphics[width=8cm]{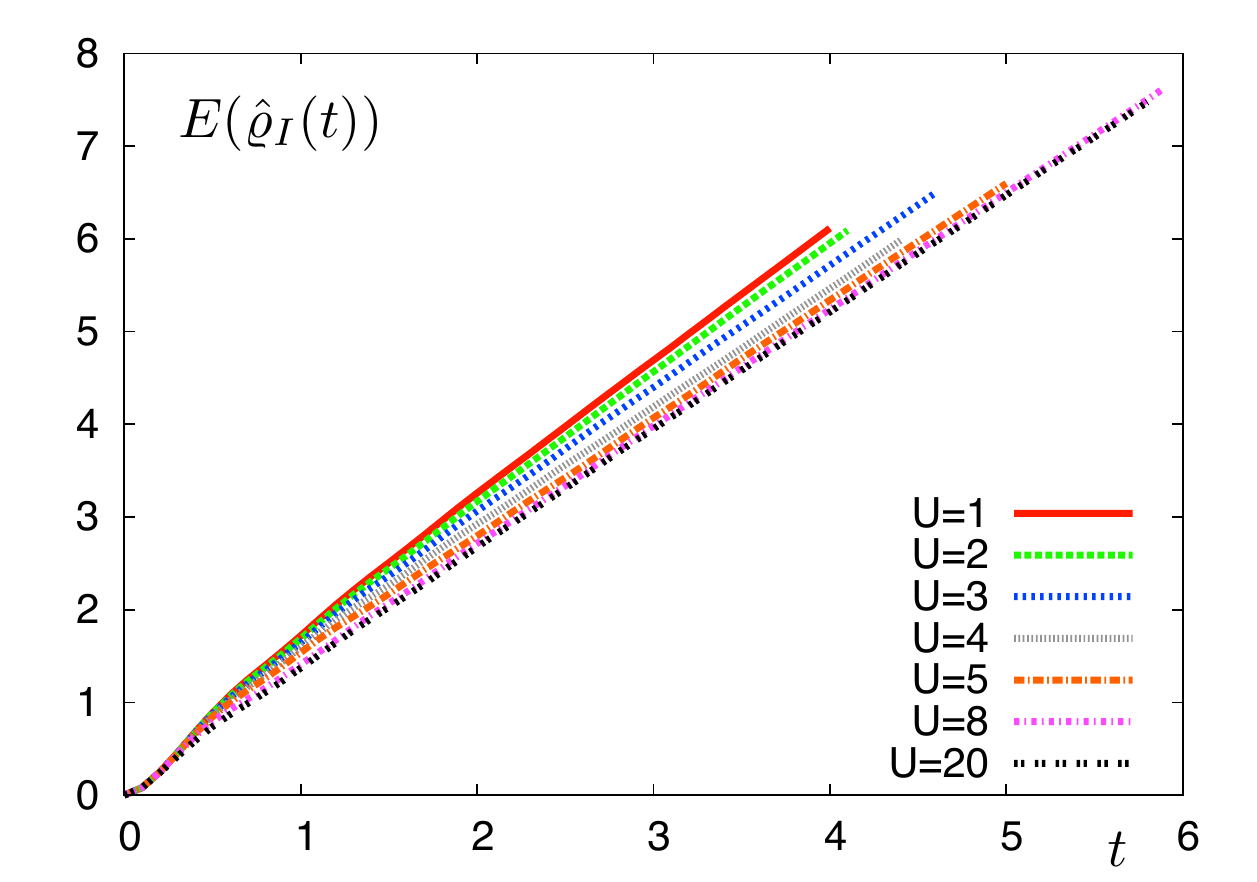}
\caption{Von-Neumann entropy of the bisected half-chain between sites
$\{1,\dots, 16\}$ and $\{17,\dots, 32\}$ vs.\ time for various interaction strengths $U$. }
\label{ent}
\end{figure}

We numerically find an initial sublinear regime in time, followed 
by a linear regime, see Fig.\ \ref{ent}. 
The linear regime is plausible when considering the linear propagation
of the excitations in the lattice. This sequence of a sublinear regime
followed by a linear one is plausible when considering the observation
that eventually, excitations travel through the lattice at a 
finite speed. This is also consistent with the above
linear upper and lower bounds. This increase also eventually limits
the time to which the time evolution can be traced using t-DMRG. 
It is interesting to observe that the entanglement growth depends only 
very weakly on interaction strength $U$, whereas the (low-energy) 
speed of sound depends quite strongly on $U$ \cite{KollathDensity}. 
This reflects the fact that after the quench, we are dealing with a 
very high energy state of the new Hamiltonian, 
where excitations are expected to be essentially ballistic 
with a propagation velocity $\propto J$, such that the
conventional low-energy speed of sound is not relevant. 

\section{Relationship to kinematical approaches}

We only very briefly touch this issue here. In this work,
similarly to Refs.\ \cite{Relax,Tegmark,Barthel}, we have considered
local relaxation to an apparent equilibrium state. The overall
system undergoes time evolution under a local Hamiltonian 
$\hat H$, whereas subsystems $I$ appear relaxed.
This observation is in an interesting relationship to kinematical approaches
\cite{Jaynes,Page,Page2,Kine1}.
Indeed, if one draws a random pure state of a large system,
this will be---in the limit of a large environment---almost always
maximally mixed. Random here means, drawn from the
unitarily invariant measure, so the Haar measure 
\cite{Page,Page2,Kine1,Kine2} (or a 
measure on the energy shell for Gaussian states \cite{Kine3}).
So one could argue that ``most states are locally relaxed''. 
This might render a relaxation to high entropy states plausible.
Yet, ironically, the image of the positive 
time axis under the time evolution corresponds, 
of course, to a one-dimensional
manifold in state space: So in order to show that 
relaxation follows from such a kinematical argument, one
has to demonstrate that for a given 
local Hamiltonian, time evolution ensures that the state of
the global system stays within the typical subspace. This is 
an interesting programme and the proof constitutes an interesting
challenge in case of interacting systems, 
which has not yet been completed. In case of free systems as in
Ref.\ \cite{Relax}, the findings may indeed be interpreted in this way.

\section{Summary}

In this work, we have introduced a setting in which 
local relaxation in quantum many-body systems
can be probed, without the need of actually
addressing single sites: This is the setting of a 
Bose-Hubbard model, in which preparation and read-out
can be done with period-$2$ translational invariance, 
as can be achieved by exploiting
optical superlattices. We have approached this idea
by deriving 
analytical expressions for the relevant quantities in 
case of the free limits of $U=0$ 
and $U=\infty$ in the Bose-Hubbard model, as well as
using  a systematic t-DMRG approach in the time-dependent
interacting case. For the $U=0$ case, we
presented  in detail 
a true convergence of subsystems to the 
maximum entropy state in $1$-norm.
In several ways, the interacting setting
reminded of the non-interacting case, certainly concerning
the mechanism of relaxation. The time-scales are different,
however, showing a faster relaxation compared to the 
inverse square-root dependence. Also, the dependence on
the interaction strength reflects the same dependence in 
the corresponding Gibbs state of the Hamiltonian the 
system is quenched to.

In this way, signatures of local relaxation
can be measured with present technology: Local densities
are found to 
relax to those of
a quasi-thermal state on well-defined time scales,
often with a similar to inverse square-root time dependence.
More sophisticated measurements would reveal
correlators and density-density correlations, again showing
local relaxation in a characteristic fashion. Hence, by
exploiting period-$2$ translational invariance, one has a 
tool at hand that effectively can be viewed as if one looked
at a small local subsystem, showing all signatures of the
local relaxation. In turn, the quasi-momentum distribution 
is a global quantity, one that shows in the free limit of $U=0$ no 
relaxation at all, and merely detects the boundary conditions
quickly. 

Hence, this is a quite exciting situation that both the
presence of the memory of the initial condition could be 
experimentally probed---in the absence of relaxation in 
``too global quantities''---as well as local relaxation: Locally,
the system ``appears relaxed'', for very long times, until
recurrences become relevant. The
technology offered by recent advances in the use of optical
superlattices should hence open up the way to experimentally
access such fundamental and old questions as the mechanism
of local relaxation in quantum many-body systems. 

\section{Acknowledgements}

We would like to thank 
T.\ Barthel, 
P.\ Calabrese, 
R.\ Fazio, 
M.B.\ Hastings, 
C.\ Kollath,
T.J.\ Osborne, 
B.\ Paredes, 
and M.M.\ Wolf for  
discussions related to the subject of local relaxation in quenched
quantum many-body systems. We specifically thank 
I.\ Bloch for several fruitful discussions on non-equilibrium
dynamics and optical superlattices, and for the 
kind hospitality in Mainz. This work has been supported
by the EU (QAP, COMPAS), the DFG (FOR 801),
the EPSRC, the EURYI, and Microsoft Research.

\section{Appendix: Proofs}
We largely follow Ref.\ \cite{Relax} to show local relaxation.
For the special situation at hand---one 
spatial dimension and $|\psi(0)\rangle=|1,0,\dots,1,0\rangle$---
the proofs simplify significantly and we state them here for completeness.

\subsection{Preliminaries}

The bosonic operators evolve in time according to (this can be 
shown by solving Heisenberg's equation of motion or applying the 
Baker-Hausdorff formula)
\begin{equation}
\label{HP_explicit}
\hat{b}_i(t)=
\me^{\mi t \hat{H}}\hat{b}_i\me^{-\mi t \hat{H} }=\sum_{j=1}^LV_{i,j}\hat{b}_j,\;\;
V(L,t)=\me^{-\mi t H},
\end{equation}
where the entries of the Hamilton matrix $H$ are given by $H_{i,j}=-\delta_{\text{dist}(i,j),1}$. For the translationally invariant setting
one has
\begin{equation}
\label{prop_matrix}
V_{i,j}(L,t)=\frac{1}{L}\sum_{k=1}^L\me^{-\mi t\lambda_k}\me^{2\pi\mi k(i-j)/L},
\end{equation}
where 
\begin{equation}
\lambda_k=-2\cos(2\pi k/L)
\end{equation}
are the 
eigenvalues of $V$.

\subsection{Proof of local relaxation}
\label{sec:proof_relax}
In this subsection, we rigorously show that for large times and
large system size, the state of a subblock $I$ becomes exactly
a maximum entropy state. The key ingredients to this proof
are Lieb-Robinson ideas and a central limit-type theorem. The
proof is not too technically involved, 
but also far from being trivial, and we 
present it for completeness. It is, after all,
quite astonishing that one does arrive at maximum entropy
states without a time average. This proof largely follows
Ref.\ \cite{Relax}, adapted to our situation of an alternating
sequence of bosons and no bosons per site as initial
condition.

We can now no longer
merely think in terms of moments, as we want to 
prove convergence of the state itself. Instead of studying the quantum state,
we investigate its characteristic function in phase space. Pointwise convergence
of the characteristic function implies convergence in trace norm for the state.
For subsets $I\subset \mathcal{L}= \{1,\dots, L\}$, the local 
state on $I$ is given by a partial trace
\begin{eqnarray}
   \hat{\varrho}_I(L,t)  = 
   \text{tr}_{\mathcal{L}\backslash I} 
   \bigl[|\psi(t)\rangle\langle\psi(t)|\bigr],
\end{eqnarray}
and the corresponding characteristic function to represent the state 
$\hat{\varrho}_I$  is defined as
\begin{equation}\label{eq:charfun}
   \chi(\vec{\alpha};t) = \text{tr}\Bigl[
   \hat{\varrho}_I\prod_{i\in I}
   \hat{D}_i(\alpha_i) \Bigr],\;
   \hat{D}_i=\me^{\alpha_i \hat{b}_i^\dagger - \alpha_i^*\hat{b}_i}.
\end{equation}
Here the $\alpha_i\in \mathbb{C}$ are the
complex phase space coordinates and
\begin{equation}
   |\psi(t)\rangle = \me^{-\mi t \hat{H} } 
   | 1,0\dots,1,0\rangle
\end{equation}
is the time evolved state vector. 
We aim at showing that the state of any subblock $I=\{1,\dots,s\}$
of $s$ consecutive sites locally relaxes. For any such block,
we find from Eq.\ (\ref{HP_explicit}) that
\begin{equation}
\chi=\langle 1,0,\dots,1,0|\prod_{i\in \mathcal{L}}
   \hat{D}_i(\beta_i) | 1,0,\dots,1,0\rangle,
  \end{equation}
where we defined
\begin{equation}
\beta_i(t)=\sum_{j=1}^s\alpha_jV_{j,i}^*(t).
\end{equation}
We thus find the following explicit form of the characteristic function of $\hat{\varrho}_I(L,t)$:
\begin{equation}
\label{eq:char_fun}
\begin{split} 
\chi  
&=\prod_{i=1}^{L/2}\langle 0|\hat{D}_{2i}(\beta_{2i})|0\rangle\prod_{i=1}^{L/2}\langle 1|\hat{D}_{2i-1}(\beta_{2i-1})|1\rangle\\
&=\me^{-\vec{\beta}^\dagger\vec{\beta}/2}
\prod_{i=1}^{L/2}(1-|\beta_{2i-1}|^2),
\end{split}
\end{equation}
where  we made use of the explicit matrix elements of the displacement operator in the Fock basis. Now, $\vec{\beta}^\dagger\vec{\beta}=\vec{\alpha}^\dagger\vec{\alpha}$ follows from unitarity of $V$ and
we proceed by proving that the above product converges pointwise in $\vec{\alpha}$ to $\exp(-\vec{\alpha}^\dagger\vec{\alpha}/2)$.

\subsubsection{The causal cone}
The key ingredient for most of what follows is the fact that the entries $V_{i,j}=V_{i-j}$ become arbitrarily small for sufficiently large $L$ and $t$. This can be seen as follows:
The $V_{i,j}=V_{i-j}$ may be thought of as Riemann-sum approximation to the
integral
\begin{equation}
\frac{1}{2\pi}
\int_{0}^{2\pi}\md\phi\,\me^{2\mi Jt \cos(\phi)}
\me^{\mi l \phi }
=\mi^lJ_l(2Jt),
\end{equation}
where $J_l$ is a Bessel function of the first kind, for which one has $J_l(x)\le x^{-1/3}$ for all $x\ge 0$ and all $l$. The error involved in this approximation is
\begin{equation}
\left|V_l-\mi^lJ_l(2Jt)
\right|\le\frac{\pi(l+2Jt)}{L},
\end{equation}
i.e., we have the bound
\begin{equation}
\left|V_l\right|\le\frac{\pi(l+2Jt)}{L}+(2Jt)^{-1/3},
\end{equation}
which converges 
to zero if we fix $l$, let $L\rightarrow\infty$ and then $t\rightarrow\infty$. 
However, we need a bound on the entries of $V$ for all $l$. To this end we complement the above bound with a Lieb-Robinson type bound on the influence of sites with large $l$: As $H_{i,j}=0$ for $\text{dist}(i,j)>1$, we have $(H^n)_{i,j}=0$ for $n<\text{dist}(i,j)=:d$, i.e.,
\begin{equation}
V_{i,j}=\sum_{n\ge d}\frac{(\mi t)^n}{n!}\left(H^n\right)_{i,j}.
\end{equation}
Now, for any matrix $M$ one has $|M_{i,j}|\le \|M\|$, where $\|\cdot\|$ indicates the operator norm. Furthermore, $n!\ge (n/3)^n$. We thus find
\begin{equation}
\left|V_{i,j}\right|\le\sum_{n\ge d}\frac{\left(6J t\right)^n}{n^n}
\le \sum_{n\ge d}\frac{\left(6J t\right)^n}{d^n}=\frac{(6Jt/d)^{d}}{1-6Jt/d},
\end{equation}
independent of $L$. Hence, matrix entries $V_{i,j}$ with $\text{dist}(i,j)>6Jt$ are exponentially suppressed
in $\text{dist}(i,j)$, defining
a "causal cone" as the influence of sites with $\text{dist}(i,j)>6Jt$ is exponentially small in 
$\text{dist}(i,j)$. For given $\varepsilon>0$ we have $|V_{i,j}|<\varepsilon$ if $\text{dist}(i,j)>B(t)$, where $B(t)$ is given by the solution to $(6Jt/B(t))^{B(t)}=\varepsilon(1-6Jt/B(t))$.
A crude bound on $B(t)$ may be obtained from noting that
\begin{equation}
\frac{(6Jt/d)^{d}}{1-6Jt/d}\le\frac{6Jt/d}{1-6Jt/d}.
\end{equation}
Combining this with the bound for $i,j$ inside the cone from above, we find for given $\varepsilon>0$ and {\em all} $i,j$ that
\begin{equation}
|V_{i,j}|<\varepsilon \text{ for all } t:\frac{4}{\varepsilon^2}<Jt<\frac{L}{4\pi^2}\frac{\varepsilon^2}{8\varepsilon+6}.
\end{equation}
The entries of $V$ are thus arbitrarily small for sufficiently large $L$ and $t$. In particular
$\lim_{t\rightarrow\infty}\lim_{L\rightarrow\infty} |V_{i,j}|=0$.

\subsubsection{Central limit-type argument}

We have from the previous section that $|V_{i,j}|$ becomes arbitrarily small for sufficiently $L$ and $t$.
Thus, for given $\vec{\alpha}$ the $|\beta_i|\le\sum_{j=1}^s|\alpha_j||V_{j,i}|$
become arbitrarily small, in particular we may assume $|\beta_i|<1/2$. 
Now, for $|x|<1/2$ one has $|\log(1+x)-x|\le x^2$, i.e.,
\begin{equation}
\label{eq:up_low_bounds}
\begin{split}
\sum_{i=1}^{L/2}\left|\log(1-|\beta_{2i-1}|^2)+|\beta_{2i-1}|^2\right|
\le \sum_{i=1}^{L/2}|\beta_{2i-1}|^4\\
\le\sup_j|\beta_{j}|^2\sum_{i=1}^{L/2}|\beta_{2i-1}|^2.
\end{split}
\end{equation}
Now,
\begin{equation}
\sum_{i=1}^{L/2}  |\beta_{2i-1}|^2=\sum_{k,l=1}^s\alpha_k^*\alpha_l
\sum_{i=1}^{L/2}V_{k,2i-1}V_{l,2i-1}^*,
\end{equation}
where, using Eq.\ (\ref{prop_matrix}),
\begin{equation}
\begin{split}
\sum_{i=1}^{L/2}V_{k,2i-1}V_{l,2i-1}^* =
\frac{1}{L^2}\sum_{r,s=1}^L\me^{\mi t(\lambda_s-\lambda_r)}\me^{\frac{2\pi}{L}\mi (r(k+1)-s(l+1))}\\
\times\sum_{i=1}^{L/2}\me^{\frac{4\pi}{L}\mi i(s-r)},
\end{split}
\end{equation}
where we find for the last line
\begin{equation}
\begin{split}
\frac{2}{L}
\sum_{i=1}^{L/2}\me^{2\pi\mi i(s-r)/(L/2)}
=\delta_{s,r}+\delta_{s-r,L/2}+\delta_{r-s,L/2},
\end{split}
\end{equation}
i.e.,
\begin{equation}
\begin{split}
\sum_{i=1}^{L/2}V_{k,2i-1}^*V_{l,2i-1}&=\frac{1}{2}\delta_{k,l}\hspace{3cm}\\
&\;\;\;\;-\frac{(-1)^l}{2L}\sum_{r=1}^L\me^{4J\mi t\cos(2\pi r/L)}
\me^{\frac{2\pi}{L}\mi r(k-l)}\\
&=\frac{1}{2}\delta_{k,l}-\frac{(-1)^l}{2}V_{k,l}(2t).
\end{split}
\end{equation}
Thus,
\begin{equation}
	\begin{split}
	\Bigl|\sum_{i=1}^{L/2}  |\beta_{2i}|^2-\frac{\vec{\alpha}^\dagger\vec{\alpha}}{2}
	\Bigr|
	\le
	\sum_{i,j=1}^s|\alpha_k||\alpha_l|V_{k,l}(2t)|,
	\end{split}
\end{equation}
which converges to zero for large $L$ and $t$.
We thus arrive at the desired statement:
\begin{equation}
\lim_{t\rightarrow\infty}\lim_{L\rightarrow\infty}\chi(\vec{\alpha};L,t)=\me^{-\vec{\alpha}^\dagger\vec{\alpha}}
\end{equation}
pointwise in $\vec{\alpha}$.

\subsection{\label{appendix:analytics:zero}Correlations after a sudden quench ($\mathbf{U=0}$)}

In this section, we give explicit forms for the two-point correlations,
\begin{equation}
f_{i,j}=\langle\hat{b}_i^\dagger(t)\hat{b}_j(t)\rangle,
\end{equation}
and density-density correlations,
\begin{equation}
g_{i,j}=\langle\hat{n}_i(t)\hat{n}_j(t)\rangle,
\end{equation}
for the translationally invariant non-interacting case, $U=0$. We already know the asymptotic behavior from the
previous section (for long times, neighboring sites become uncorrelated as the state converges to a direct product of Gaussians) but the goal here is to derive explicit expressions for all times.

We find
from Eqs.\ (\ref{HP_explicit},\ref{prop_matrix}) that
\begin{equation}
\begin{split}
f_{i,j}&=\frac{1}{L}\sum_{p,q=1}^L\me^{\mi t(\lambda_q-\lambda_p)}\me^{2\pi\mi (pj-qi)/L}\\
&\hspace{2cm}\times\frac{1}{L}\sum_{k=1}^{L/2}\me^{2\pi\mi (2k-1)(q-p)/L},
\end{split}
\end{equation}
where the term in the last line evaluates to
\begin{equation}
\begin{split}
\frac{\me^{-2\pi\mi (q-p)/L}}{L}\sum_{k=1}^{L/2}\me^{2\pi\mi k(q-p)/(L/2)}\hspace{2cm}\\
=\frac{1}{2}\left(\delta_{p,q}-\delta_{q-p,L/2}-\delta_{p-q,L/2}\right).
\end{split}
\end{equation}
We thus have
\begin{equation}
\begin{split}
f_{i,j}=\frac{1}{2}\delta_{i,j}-\frac{(-1)^i}{2L}\sum_{p=1}^L\me^{4J\mi t\cos(2\pi p/L)}
\me^{2\pi\mi p(j-i)/L},
\end{split}
\end{equation}
which yields a particularly simple form in the thermodynamic limit $L\rightarrow\infty$
\begin{equation}
f_{i,j}\rightarrow \frac{1}{2}\delta_{i,j}-\frac{(-1)^i\mi^{j-i}}{2}J_{j-i}(4Jt),
\end{equation}
where $J_n$ is a Bessel function of the first kind.
Thus, the total number of particles at even sites
\begin{equation}
\sum_{i\text{ even}}\left\langle\hat{n}_i(t)\right\rangle=\frac{L}{4}-\frac{L}{4}J_0(4Jt)
\end{equation}
is a truly local quantity in this translationally invariant setting.

The above allows us to derive the quasi-momentum distribution,
\begin{equation}
\begin{split}
	S(q,t)&=\frac{1}{L}\sum_{i,j}\me^{\mi q(i-j)}f_{i,j}\\
	&=\frac{1}{2}-\frac{1}{2L^2}\sum_{p=1}^L\me^{4J\mi t\cos(2\pi p/L)}\\
&\;\;\;\;\;\;\times\sum_{i,j=1}^L
\me^{\mi i(q+\pi-2\pi p/L)}
\me^{\mi j(2\pi p/L-q)},
\end{split}
\end{equation}
where the last term vanishes for $Lq/(2\pi)\in \{1,\dots,L\}$, 
yielding $S(q,t)=1/2$. For all other $q\in[0,2\pi]$ we find
\begin{equation}
S(q,t)=\frac{1}{2}+\frac{\mi}{L^2}\sum_{p=1}^L\me^{4J\mi t\cos(2\pi p/L)}\frac{\sin^2(Lq/2)}{\sin(2\pi p/L-q)}.
\end{equation}

To derive the density-density correlations, we first observe that for 
\begin{equation}
	h_{k,l,r,s}=\langle n_1,n_2,\dots,n_L|\hat{b}^\dagger_k
	\hat{b}_l\hat{b}^\dagger_r\hat{b}_s|n_1,n_2,\dots,n_L\rangle, 
\end{equation}
we have 
(writing $\delta_{k\ne l}=1-\delta_{k,l}$)
\begin{equation}
\label{eq:h1}
\begin{split}
h_{k,l,r,s}&=\left(\delta_{k,l}+\delta_{k\ne l}\right)\left(\delta_{r,s}+\delta_{r\ne s}\right)h_{k,l,r,s}\\
&=\delta_{k,l}\delta_{r,s}n_kn_r
+\delta_{k\ne l}\delta_{r\ne s}h_{k,l,r,s}\\
&=\delta_{k,l}\delta_{r,s}n_kn_r+\delta_{k\ne l}\delta_{k,s}\delta_{l,r}n_k(1+n_l).
\end{split}
\end{equation}
Thus,
\begin{equation}
\begin{split}
g_{i,j}=&\sum_{k=1}^{L/2}V_{i,2k-1}^*V_{i,2k-1}\sum_{r=1}^{L/2}V_{j,2r-1}^*V_{j,2r-1}\\
&+2\sum_{k,l=1}^{L/2}\delta_{k\ne l}V_{i,2k-1}^*V_{i,2l-1}V_{j,2l-1}^*V_{j,2k-1},
\end{split}
\end{equation}
and therefore,
\begin{equation}
\begin{split}
g_{i,j}=&f_{i,i}f_{j,j}+f_{i,j}(\delta_{i,j}+f_{j,i})\\
&-2\sum_{k=1}^{L/2}|V_{i,2k-1}|^2|V_{j,2k-1}|^2.
\end{split}
\end{equation}
This expression yields
\begin{equation}
\begin{split}
&\hspace{-1.5cm}\langle\hat{N}_e(t)\hat{N}_o(t)\rangle-\langle\hat{N}_e(t)\rangle\langle\hat{N}_o(t)\rangle\\
=&\sum_{i,j=1}^{L/2}\left(g_{2i,2j-1}-f_{2i,2i}f_{2j-1,2j-1}\right)\\
=&\sum_{i,j=1}^{L/2}f_{2i,2j-1}f_{2j-1,2i}\\
&-2\sum_{k=1}^{L/2}\sum_{i,j=1}^{L/2}|V_{2i,2k-1}|^2|V_{2j-1,2k-1}|^2,
\end{split}
\end{equation}
where, using the unitarity of $V$,
\begin{equation}
\begin{split}
&\hspace{-0.5cm}\sum_{i,j=1}^{L/2}|V_{2i,2k-1}|^2|V_{2j-1,2k-1}|^2\\
&=\left(1-\sum_{i=1}^{L/2}|V_{2i-1,2k-1}|^2\right)\sum_{j=1}^{L/2}|V_{2j-1,2k-1}|^2\\
&=(1-f_{2k-1,2k-1})f_{2k-1,2k-1},
\end{split}
\end{equation}
and, using the explicit  form of $f_{i,j}$ and identifying Kronecker deltas,
\begin{equation}
\begin{split}
&\sum_{i,j=1}^{L/2}f_{2i,2j-1}f_{2j-1,2i}\\
&=-\frac{1}{16}\sum_{\substack{p,q=1\\ \frac{p+q}{L/2}\in\zz}}^L
\me^{4J\mi t(\cos(2\pi p/L)+\cos(2\pi q/L))-2\pi\mi(p-q)/L}\\
&=-\frac{1}{16}\sum_{p=1}^L
\me^{8J\mi t\cos(2\pi p/L)}+\frac{L}{16}.
\end{split}
\end{equation}
For large $L$ we then find
\begin{equation}
\begin{split}
&\langle\hat{N}_e(t)\hat{N}_o(t)\rangle-\langle\hat{N}_e(t)\rangle\langle\hat{N}_o(t)\rangle\\
&\hspace{0.4cm}\rightarrow
-\frac{L}{16}\left(
3+
J_0(8Jt)
-4[J_0(4Jt)]^2\right).
\end{split}
\end{equation}
\subsection{\label{appendix:analytics:infinity}Correlations after a sudden quench ($\mathbf{U\rightarrow\infty}$)}

Using the results for $U=0$ (\ref{appendix:analytics:zero}) we can now easily derive analytical expressions for local densities
\begin{equation}
\label{eq:fd_infty}
f_{i,i} = \langle \hat{n}_i(t) \rangle,
\end{equation}
next-neighbour two-point correlations
\begin{equation}
\label{eq:fnn_infty}
f_{i+1,i} = \langle  \hat{b}^\dagger_{i+1}(t) \hat{b}_i(t)\rangle
\end{equation}
and density-density correlations
\begin{equation}
\label{eq:g_infty}
g_{i,j} = \langle \hat{n}_{i}(t)\hat{n}_j(t)\rangle
\end{equation}
for $U\rightarrow\infty$. For simplicity we assume even particle numbers $N=L/2$. With Eqs.\ (\ref{eq:jordan}), (\ref{eq:state_trans}) it follows that Eqs.\ (\ref{eq:fd_infty})-(\ref{eq:g_infty}) are also valid for after the mapping to fermions, with the bosonic creation- and annihilation operators replaced by fermionic ones. Because of the identical time evolution of the bosonic- and fermionic operators (equation (\ref{eq:evolution_boson}) respectively (\ref{eq:evolution_fermion})) we find essentially the same results for the local densities and the next-neighbour two-point correlations like in section \ref{appendix:analytics:zero} as we did not make any use of commutation relations in the derivation of the bosonic respectively fermionic results:
\begin{eqnarray}
	f_{i,i}&=&
	\frac{1}{2}-\frac{(-1)^i}{2L}
	\sum_{p=1}^L\me^{4J\mi t\cos(2\pi p/L)},\\
	f_{i+1,i}&=&
	-\frac{(-1)^{i+1}}{2L}\sum_{p=1}^L\me^{4J\mi t\cos(2\pi p/L)}
	\me^{-2\pi\mi p/L}.
\end{eqnarray}
In contrast to this, the results for the density-density correlations differ. For fermionic operators a similar derivation to Eq.\ (\ref{eq:h1}) yields
\begin{equation}
\begin{split}
\tilde{h}_{k,l,r,s} =& \langle n_1,n_2,\dots,n_L \lvert \hat{c}^\dagger_k \hat{c}_l \hat{c}^\dagger_r \hat{c}_s \rvert  n_1,n_2,\dots,n_L \rangle	\\
=& \delta_{k,l}\delta_{r,s}n_kn_r+\delta_{k\ne l}\delta_{k,s}\delta_{l,r}n_k(1-n_l).
\end{split}
\end{equation}
This finally leads to
\begin{equation}
\begin{split}
g_{i,j} = \tilde{f}_{i,i}\tilde{f}_{j,j}+\tilde{f}_{i,j}(\delta_{i,j}-\tilde{f}_{j,i})
\end{split}
\end{equation}
with $\tilde{f}_{i,j} = \langle  \hat{c}^\dagger_i(t) \hat{c}_j(t) \rangle$. Using the explicit form of the time-evolved operators in Eqs.\ (\ref{eq:evolution_boson}), (\ref{eq:evolution_fermion}) and the absence of any signs resulting from commutation relations it is easy to see that $\tilde{f}_{i,j}$ is identical with  $f_{i,j}$ for free bosons (see \ref{appendix:analytics:zero}). This finally leads to Eq. (\ref{eq:dd_infty}). Using the results of \ref{appendix:analytics:zero}, the global density-density correlator is then given by
\begin{equation}
\begin{split}
&\langle\hat{N}_e(t)\hat{N}_o(t)\rangle-\langle\hat{N}_e(t)\rangle\langle\hat{N}_o(t)\rangle\\
&=-\sum_{i,j=1}^{L/2}f_{2i,2j-1}f_{2j-1,2i}\\
&=\frac{L}{16}\left(\frac{1}{L}\sum_{p=1}^L \me^{8J\mi t \cos(2\pi p/L)} - 1\right).
\end{split}
\end{equation}
For large $L$ this yields
\begin{equation}
\begin{split}
&\langle\hat{N}_e(t)\hat{N}_o(t)\rangle-\langle\hat{N}_e(t)\rangle\langle\hat{N}_o(t)\rangle\\
&\rightarrow - \frac{L}{16} \left(1-J_0(8Jt)\right).
\end{split}
\end{equation}


\begin{thebibliography}{99}
\bibitem{Momentum}
	 Similar statements may also be true in momentum
	space for some observables, and in fact, using an
	approximate Wegner flow method, such a relaxation has
	been identified in interacting systems \cite{Moeckel}.
\bibitem{LRC}
	Such bounds to the speed of information transfer
	have rigorously been 
	shown for finite-dimensional systems
	such as spins or fermions \cite{LR,Hastings,LRReview}, 
	and for free bosonic 
	systems \cite{Nachtergaele,HarmonicLiebRobinson,Buer}.
\bibitem{Note}
	Note that for long-range correlation functions, results
	from conformal field theory also strongly suggest a 
	relaxation in the asymptotic behavior 	
	when quenched to critical quantum
	systems \cite{Calabrese,CalabreseLong}.
\end{thebibliography}
\end{document}